\documentclass[journal]{IEEEtran}

\ifCLASSINFOpdf
\else
\fi
\usepackage{dblfloatfix}    

\usepackage{cite}
\usepackage{amsmath,amssymb,amsfonts}
\usepackage{bbm}

\usepackage{graphicx}
\usepackage{epstopdf}
\usepackage{textcomp}
\usepackage{xcolor}
\usepackage{colortbl} 

\usepackage{url} 

\usepackage{bm}
\usepackage[normalem]{ulem} 
\usepackage{supertabular}
\usepackage{amsthm}
\theoremstyle{definition}

\newtheorem{remark}{Remark}
\usepackage[linesnumbered,ruled,vlined]{algorithm2e}
\usepackage{algorithmicx} 
\usepackage{pbox}
\usepackage{float} 
\usepackage{subcaption} 

\usepackage{boldline,multirow}
\usepackage{array}

\usepackage{pbox}

\usepackage{dblfloatfix}

\newcounter{probNum}

\begin{document}
%
%
\title{Downlink Resource Allocation in Multiuser Cell-free MIMO Networks with User-centric Clustering}
%
%
\author{Hussein~A.~Ammar\IEEEauthorrefmark{1},~\IEEEmembership{Student Member,~IEEE}, 
	Raviraj~Adve\IEEEauthorrefmark{1},~\IEEEmembership{Fellow,~IEEE},
	Shahram~Shahbazpanahi\IEEEauthorrefmark{2}\IEEEauthorrefmark{1},~\IEEEmembership{Senior Member,~IEEE},
	Gary~Boudreau\IEEEauthorrefmark{3},~\IEEEmembership{Senior Member,~IEEE}, 
	and~Kothapalli~Venkata~Srinivas\IEEEauthorrefmark{3},~\IEEEmembership{Member,~IEEE}
	\thanks{
		\IEEEauthorrefmark{1}H. A. Ammar and R. Adve are with the Edward S. Rogers Sr. Department of Electrical and Computer Engineering, University of Toronto, Toronto, ON M5S 3G4, Canada (e-mail: ammarhus@ece.utoronto.ca; rsadve@comm.utoronto.ca).
	}
	\thanks{
		\IEEEauthorrefmark{2}S. Shahbazpanahi is with the Department of Electrical, Computer, and Software Engineering, University of Ontario Institute of Technology, Oshawa, ON L1H 7K4, Canada. He also holds a Status-Only position with the Edward S. Rogers Sr. Department of Electrical and Computer Engineering, University of Toronto.
	}
	\thanks{
		\IEEEauthorrefmark{3}G. Boudreau and K. V. Srinivas are with Ericsson Canada, Ottawa, ON K2K 2V6, Canada.
	}
	\thanks{This work was supported in part by Ericsson Canada and in part by the Natural Sciences and Engineering Research Council (NSERC) of Canada.}
	\thanks{This work was published in part in the IEEE International Conference on Communications (ICC)~\cite{ammarRA_NonCoherConf}.}
}


%
%

\maketitle 

\makeatletter
\def\tagform@#1{\maketag@@@{\normalsize(#1)\@@italiccorr}}
\makeatother

\newcommand*{\myResultReuseFactorPerfCoher}{$50.6$} 
\newcommand*{\myResultReuseFactorPerfTwoCoher}{$50.1$} 
\newcommand*{\myResultReuseFactorPerfThreeCoher}{$55.2$}
\newcommand*{\myResultReuseFactorPerfCoherNF}{$27.4$}
\newcommand*{\myResultReuseFactorPerfTwoCoherNF}{$41.2$}
\newcommand*{\myResultReuseFactorPerfThreeCoherNF}{$46.3$}

\newcommand*{\myResultOne}{$9.1$}
\newcommand*{\myResultTwo}{$10.6$}
\newcommand*{\myResultThree}{$1.67$}
\newcommand*{\myResultFour}{$5.44$}
\newcommand*{\myResultReuseFactor}{$0.32$} 
\newcommand*{\myResultReuseFactorTwo}{$0.16$} 
\newcommand*{\myResultReuseFactorThree}{$0.08$} 
\newcommand*{\myResultReuseFactorPerf}{$65.64$} 
\newcommand*{\myResultReuseFactorPerfTwo}{$70.8$} 
\newcommand*{\myResultReuseFactorPerfThree}{$78.83$} 
\newcommand*{\myResultReuseFactorPerfNF}{$49.47$} 
\newcommand*{\myResultReuseFactorPerfNFTwo}{$65.24$} 
\newcommand*{\myResultReuseFactorPerfNFThree}{$77$} 
\newcommand*{\myResultReuseFactorPerfRobust}{$39.21$} 
\newcommand*{\myResultReuseFactorPerfRobustTwo}{$37$} 
\newcommand*{\myResultReuseFactorPerfRobustThree}{$43.21$}
\newcommand*{\myResultReuseFactorPerfRobustNF}{$10.6$} 
\newcommand*{\myResultReuseFactorPerfRobustNFTwo}{$25$} 
\newcommand*{\myResultReuseFactorPerfRobustNFThree}{$38.27$}
\newcommand*{\myResultReuseFactorPerfRobustSingleTS}{$37$} 
\newcommand*{\myResultReuseFactorPerfRobustSingleTSTwo}{$28.62$} 
\newcommand*{\myResultReuseFactorPerfRobustSingleTSThree}{$32.1$} 
\newcommand*{\myResultReuseFactorPerfRobustSingleTSNF}{$7.32$} 
\newcommand*{\myResultReuseFactorPerfRobustSingleTSNFTwo}{$15.02$} 
\newcommand*{\myResultReuseFactorPerfRobustSingleTSNFThree}{$26.18$} 

\begin{abstract}
	In this paper, we optimize user scheduling, power allocation and beamforming in distributed multiple-input multiple-output (MIMO) networks implementing user-centric clustering. We study both the coherent and non-coherent transmission modes, formulating a weighted sum rate maximization problem for each; finding the optimal solution to these problems is known to be NP-hard. We use tools from fractional programming, block coordinate descent, and compressive sensing to construct an algorithm that optimizes the beamforming weights and user scheduling and converges in a smooth non-decreasing pattern. Channel state information (CSI) being crucial for optimization, we highlight the importance of employing a low-overhead pilot assignment policy for scheduling problems. In this regard, we use a variant of hierarchical agglomerative clustering, which provides a suboptimal, but feasible, pilot assignment scheme; for our cell-free case, we formulate an \emph{area-based} pilot reuse factor. Our results show that our scheme provides large gains in the long-term network sum spectral efficiency compared to benchmark schemes such as zero-forcing and conjugate beamforming (with round-robin scheduling) respectively. Furthermore, the results show the superiority of coherent transmission compared to the non-coherent mode under ideal and imperfect CSI for the area-based pilot-reuse~factors we consider.
\end{abstract}

%
\IEEEpeerreviewmaketitle

\section{Introduction}
In conventional networks, the performance of cell-edge users has been a major concern since, due to comparable received signal and interference levels, these users experience the worst network performance. In this regard, cell-free distributed multiple-input multiple-output (MIMO) has emerged as a promising concept to eliminate cell boundaries and provide reliable service for all users. Such MIMO networks, comprising distributed remote radio heads (RRHs) coordinate their operations to serve users most effectively. Cell-free networks have shown to outperform traditional networks under various performance metrics~\cite{cellFreeVersusSmallCells7827017, powerControlCellFree7917284} . For example, the study in~\cite{cellFreeVersusSmallCells7827017} reports five-fold and ten-fold improvements in achievable rate over a small-cell scheme with uncorrelated and correlated shadow fading, respectively.

Clustering is an essential component of any practical cell-free scheme. Serving all users with all transmitters in a large region is impractical; this is mainly due to the capacity of individual RRHs to serve only a limited number of users. Additionally, serving users with distant RRHs occupies resources but contributes little useful signal power. A practical scheme to deploy a cell-free network that is receiving an increasing attention is user-centric clustering. In this scheme, the concept of a cell-edge user is completely eliminated by positioning the user at the center of its serving cluster, i.e., a serving cluster of cooperating RRHs is constructed separately for each user. This formation of clusters can be based on criteria such as serving distance~\cite{PDPUsercentricVsDisjoint} or network performance~\cite{clustersbasedonNetperf7105966}. In addition to its role in practical deployments, by enabling macro-diversity and making efficient use of limited power/bandwidth resources, user-centric clustering can outperform a general cell-free network that assumes all RRHs serve all users~\cite{cellFreeUserCentricPower8901451}. 

There have been two main transmission schemes considered for such distributed networks: coherent and non-coherent transmissions. When using coherent transmissions, all RRHs coherently transmit the same data symbol to users. While, nominally, coherent transmissions require strict phase synchronization across all RRHs serving a user~\cite{CoherentCRAN8606433}, in practice, networks generally use multi-carrier transmissions and can relax this requirement as long as the relative time delays are within the chosen cyclic prefix. The implication, as our recent work~\cite{PDPUsercentricVsDisjoint} has shown, is that distributed networks without synchronization require an extended prefix; however, amongst clustering schemes, user-centric clustering suffers the least in this regard. In this paper, we do not consider synchronization issues and focus on a single subcarrier in a multi-carrier system. Specifically, in dealing with the coherent case, we will assume synchronization. An alternative approach is non-coherent transmissions where each RRH transmits a different data stream~\cite{NoncoherentCRAN8482453}. While phase synchronization is not required, there is a significant penalty in terms of rate (as we shall see) and complexity, since the user must implement successive interference cancellation (SIC) to decode the individual streams. In this work, we analyze both approaches.

This paper focuses on resource allocation in distributed MIMO, but not massive MIMO, networks where a set of RRHs need to serve a much larger set of users. Nominal serving clusters are formed using a threshold on average channel power. However, given the limited number of users each RRH can serve, user scheduling remains to be determined. In this regard, we develop an algorithm to perform user scheduling and beamforming (implicitly including power allocation). Specifically, we formulate a weighted sum rate (WSR) maximization problem; we note that WSR problems have been shown to be NP-hard in conventional networks~\cite{Complexity4453890}. 

Recently, resource allocation for cell-free MIMO networks has received significant attention. The investigations in~\cite{PrecodingDistrib2020Atzeni, 6151868} optimize the beamforming by minimizing the weighted sum mean square error (MSE). Similarly, in~\cite{6920005} the authors study beamforming design converting the problem into an MSE-minimization problem. We note that these works minimize the weighted MSE, which is easier to solve than a WSR problem, but suffer a penalty in terms of sum-rate performance~\cite{PrecodingDistrib2020Atzeni}. The work in~\cite{differentCooperationLevels8845768} focuses on beamforming under different levels of cooperation between the RRHs and their control unit (CU). On the other hand, in~\cite{9064545} the authors study dynamic clustering while focusing on estimation of channel state information (CSI).

The work in~\cite{cellFreeUserCentricPower8901451} performs power allocation by maximizing a lower-bound on sum rate or on the minimum rate. Similarly, the work in~\cite{maxMinRate8756286} considers the max-min problem to optimize beamforming. We note that, in fairness sense, max-min approaches are an extreme with all users achieving the same rate (if feasible). On the  other hand, WSR provides a flexible framework to enable fairness; in this paper we use the weights to impose proportional fairness (PF). The authors in~\cite{powerControlCellFree7917284} consider  near-optimal power control using zero-forcing (ZF) and conjugate beamforming; this is simpler than solving for the max–min power for cell-free massive MIMO networks. The work in~\cite{9136914, 8970501} performs power allocation, switching off some RRHs to foster energy efficiency. Furthermore, in~\cite{pilotPowerControl8450041} the authors control the power to decrease pilot contamination. Also,~\cite{EnergyE2020towards, EnergyEfficiency8097026, 9136914}, study energy efficiency under a cell-free distributed MIMO scheme.

The study in~\cite{6815733} optimizes user scheduling for a cluster of coordinated base stations (BSs), with a constraint on the maximum transmit power. However, here the users are connected to only one BS. Moreover, the work in~\cite{Cell-freeWSR2005.12331} uses the WSR to optimize the beamforming by approximating the problem by a conic-quadratic program based on the inner approximation framework~\cite{marks1978general}, where the authors use a lower-bound for the logarithm function to obtain a local optimum. The authors also report the globally optimal solution using a branch-reduce-and-bound framework (unfortunately, with exponential complexity~\cite{tuy2005monotonic}). Furthermore, the work does not consider user scheduling, nor does it provide closed-form expressions for the required beamforming weights.

In addition to studies that focus on cell-free networks, we review some innovative techniques specifically targeting user scheduling for conventional networks. The work in~\cite{ResourceAllo6175089} uses dual-decomposition and gradient methods for resource allocation in a conventional MIMO relaying system, where the binary scheduling variables, interpreted as time-sharing factors, are initially relaxed. Notably, this relaxation still leads to binary solutions for a large number of subcarriers~\cite{ResourceAllo1658226}. The work in~\cite{ResourceAllo793310} applies the same procedure to perform subcarrier, bit, and power allocation to minimize the total transmit power while satisfying a given transmission rate threshold.

As an alternative approach, the studies in~\cite{FR8310563, Ahmad9084256} use fractional programming to perform resource allocation and user scheduling in conventional networks, where user scheduling part is performed by matching the non-zero beams resulting from each algorithm iteration to the users through a combinatorial search. The work in~\cite{DistribResourceAllo7676375, FullyDistResourceAllo6630117} uses a game-theoretic approach to schedule users in a conventional network using a distributed form of auction theory; the users compete for the resources through bidding and assignment phases~\cite{bertsekas1979distributed}. 
One disadvantage of such an approach is the large communication overhead between the different network entities before convergence (an allocation) occurs. In addition, these studies rely on a carrier sense multiple access scheme, which allows the high bidders to first transmit by relating their bid value to the inverse of their transmit backoff timer. These schemes are, therefore, not suitable for MIMO systems. Moreover, such schemes do not guarantee strong optimality~\cite{DistribResourceAllo7676375}.


In summary, the main limitations of these works are: first, most do not specifically address the user-centric clustering scheme at hand. Second, most ignore user scheduling and assume that the users have been \emph{pre-selected}, e.g., randomly or through a round-robin selection, neither of which provides rate-fairness. Third, most of these schemes do not provide closed form expressions for the optimization variables, which precludes any analysis and/or does not guarantee convergence.

In this paper, we analyze the downlink of a distributed MIMO network using user-centric clustering. Specifically, the contributions of this paper are:
\begin{itemize}
	\item We optimize resource allocation, in terms of user scheduling and beamforming, to maximize the WSR for both the coherent and non-coherent transmission modes. We note that the needed user scheduling cannot be performed efficiently using combinatorial algorithms because the utility function depends on the signals received from many RRHs.
	\item We obtain closed-form relations between the optimization variables, leading to an iterative algorithm that converges smoothly.
	\item We highlight the problem of non-orthogonal pilots assignment (PA) when the users to be scheduled are still not determined. To do so, we introduce an \emph{area-based} pilot reuse factor appropriate for distributed networks. Specifically, we propose a low-overhead PA policy and quantify the performance loss due to imperfectly estimated CSI, where we use robust beamforming to decrease this loss. 
	\item We present numerical results to illustrate the substantial gains over benchmark schemes such as ZF and conjugate beamforming with round-robin scheduling.	
\end{itemize}

\begin{figure*}[t]
	\vspace{-1.7em}
	\centering
	\includegraphics[width=0.6\linewidth]{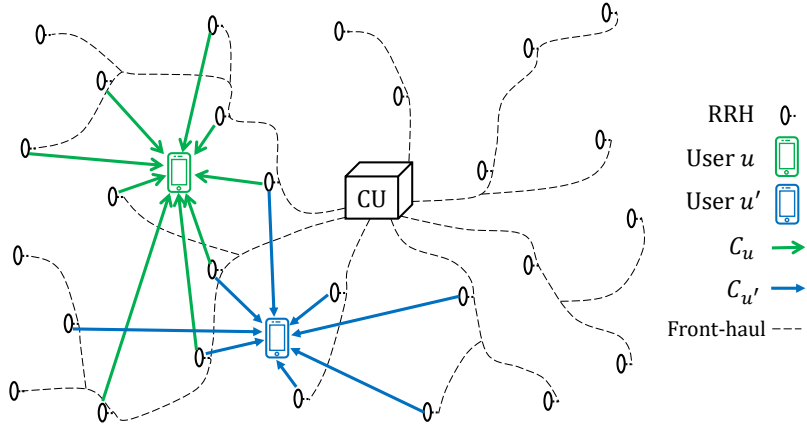}
	\vspace{-1em}
	\caption{Serving cluster using user-centric clustering.}
	\label{fig:UserCentricClustering}
	\vspace{-1.5em}
\end{figure*}

The rest of the paper is organized as follows. Section~\ref{section:Model} presents the system model, while Section~\ref{section:Formulation} defines our optimization problem under coherent transmission. Section~\ref{section:Formulation} provides a detailed analysis, where we introduce the steps necessary to solve the problem efficiently. In Section~\ref{section:ResourceAllocation}, we present our resource allocation optimization algorithm alongside closed-form expressions to update the optimized variables using coordinate descent; we also treat the perfect CSI case as a special case. 
Section~\ref{sec:NonCoher} tackles the problem for the non-coherent transmission mode. Section~\ref{section:results} reports on our simulation results illustrating the gains possible by effective resource allocation. Finally, Section~\ref{section:conclusion} concludes the paper.

\emph{Notation:} 
Both lower and upper case letters (e.g., $a$ and $A$) represent scalars, while their bold counterparts ${\bf a}$ and ${\bf A}$ represent vectors and matrices, respectively. Operators $\mathtt{tr}(\cdot)$, $(\cdot)^{-1}$, $(\cdot)^T$, $(\cdot)^*$, and $(\cdot)^H$, denote the trace, inverse, transpose, conjugate, and conjugate transpose, respectively. $\mathbb{E}\{\cdot\}$ represents statistical expectation, $\|\cdot\|_2$ and $|\cdot|$ are the vector and scalar Euclidean norms, $\|\cdot\|_p$ is the $\mathrm{l}_p$-norm, ${\bf A} = {\rm diag} \left({\bf a}\right)$ is a diagonal matrix with diagonal constructed from vector ${\bf a}$, ${\rm vec}\left(\cdot\right)$ is the vectorization operator, $\otimes$ is the Kronecker product, and ${\bf I}_m$ is $m \times m$ identity matrix. For a set $\mathcal{A}$, $|\mathcal{A}|$ denotes its cardinality. Finally, $\mathbb{B}$, $\mathbb{R}$, $\mathbb{C}^{m\times 1}$, and $\mathbb{C}^{m\times n}$ represent binary numbers, real numbers, complex $m\times 1$ vectors, and complex $m\times n$ matrices, respectively.

\section{System Model}\label{section:Model}
\subsection{Network Model}
We consider a wireless network as shown in Fig.~\ref{fig:UserCentricClustering}. Users, represented by the set $\mathcal{U}$, are served by RRHs, represented by the set $\mathcal{B}$, both distributed in the network. The RRHs, each equipped with $M$ antennas, cooperate among themselves to serve the active users. Unlike in the massive MIMO setting, $M$ is a relatively small number. As with other studies, e.g.,~\cite{cellFreeUserCentricPower8901451}, we assume that the RRHs are connected to a CU through a perfect front-haul network, e.g., using wired front-haul technologies like fiber optics or the radio stripes system~\cite{frenger2019antenna}. This assumption allows us to focus entirely on the resource allocation problem at hand.

Based on user-centric clustering, for each user $u \in \mathcal{U}$, we define a set, or \textit{serving cluster}, $\mathcal{C}_u \subset \mathcal{B}$, comprising the RRHs that \emph{potentially} serve this user. The set comprises the RRHs $r$ that provide an average channel gain to user $u$ above a chosen threshold, i.e., $\left(\psi_{ru}\ell(d_{ru}) \right) \ge \rho$, where $\psi_{ru}$ accounts for the shadowing, $\ell(d_{ru})$ accounts for the path loss and $d_{ru}$ is the corresponding distance. As can be seen in Fig.~\ref{fig:UserCentricClustering}, because of shadowing, an RRH relatively close to a user may not be in its serving cluster. If, for a specific user, no RRH can meet this criterion, its corresponding cluster includes the single RRH providing the largest average channel gain, i.e., largest $\{\left(\psi_{ru}\ell(d_{ru}) \right) : r \in \mathcal{B}\}$. Mathematically, this can be represented as $\mathcal{C}_u = \{r: \left(\psi_{ru}\ell(d_{ru}) \right) \ge \rho\} \cup \{{\rm arg}\ \max_r\ \psi_{ru}\ell(d_{ru})\}$. Defining a cluster for each user $u$ is useful to exclude the RRHs that cannot significantly contribute to the user's signal, limiting the required signaling overhead. 

Unlike several studies in the literature, we do not assume that the users are already scheduled; indeed, user-scheduling is a key step in our optimization. Importantly, we do not assume a massive MIMO scenario where the number of serving antennas found in the network far exceeds the number of users $|\mathcal{U}|$, i.e., not all users can be scheduled simultaneously. Our model is based on two facts: first, the number of deployed RRHs is much smaller than the number of users, at least in the near future, and this is coupled by the small number of antennas per RRH to allow for low-cost RRHs; second, the number of users (e.g., mobile phones, tablet, sensors, nodes in an internet of things) is increasing at a rate faster than the deployment density of RRHs. 
	
In our cell-free scheme, we expect only a few RRHs will serve each user. Thus, we require a scheme to schedule the users and allocate the resources. Below, we choose fairness weights to set users priorities and prevent the repeated scheduling of the users with best channel in every time slot.

\subsection{Signal Model for Coherent Transmission}
For every RRH $r \in \mathcal{B}$, we define the set $\mathcal{E}_{r}$ representing the users to be served by that RRH. These sets can be directly obtained from the sets $\left\{\mathcal{C}_u: u \in \mathcal{U}\right\}$, where $u \in \mathcal{E}_r \Leftrightarrow r \in \mathcal{C}_u$. In a particular time slot, not all the users are necessarily served because the number of users is much greater than the available spatial resources; deciding which users are to be served is determined by our user scheduling algorithm. Importantly, as can be seen in Fig.~\ref{fig:UserCentricClustering}, unlike in the case of disjoint clustering, with user-centric clustering, any set $\mathcal{E}_{r}$ may only partially overlap with another set $\mathcal{E}_{r'}$, for $r' \ne r$ i.e., each set $\mathcal{E}_r$ may be~unique.

In this paper we focus on a single narrowband transmission; this may represent, for example, a single sub-carrier in a multi-carrier transmission. In the coherent transmission mode, for each symbol period, the signal model at user $u$ with serving cluster $\mathcal{C}_u$ can be written as
\begin{align}\label{eq:signalModel_Coherent}
y_{u} &=
\sum_{r \in \mathcal{C}_u} \sqrt{{s}_{ru}}
\left({\bf \bar{h}}_{ru}\right)^H {\bf \bar{w}}_{ru} x_{u}
\nonumber \\
&\ + \sum_{u' \in \mathcal{U}_{-u}} \sum_{r' \in \mathcal{C}_{u'}} \sqrt{{s}_{r'u'}} \left({\bf \bar{h}}_{r'u}\right)^H {\bf \bar{w}}_{r'u'} x_{u'}
+ z_{u}
\nonumber \\
& =
\underbrace{{\bf h}_{u,u}^H {\bf S}_{u}^{1/2} {\bf w}_{u} x_{u}}_{\mathrm{signal}}
+ \underbrace{\sum_{u' \in \mathcal{U}_{-u}} {\bf h}_{u',u}^H {\bf S}_{u'}^{1/2} {\bf w}_{u'} x_{u'}}_{\mathrm{multiuser~interference}}
+ \underbrace{z_{u}}_{\mathrm{noise}},
\end{align}
where $\mathcal{U}_{-u} = \{\mathcal{U} \backslash u \}$, and $x_{u}$ denotes the data symbol transmitted to user $u$ by its serving cluster, with $\mathbb{E}\{|x_{u}|^2\} = 1$. Furthermore,  ${\bf \bar{h}}_{ru} \triangleq \sqrt{\psi_{ru} \ell(d_{ru})} {\bf g}_{ru} \in \mathbb{C}^{M \times 1}$ denotes the channel between RRH $r$ and user $u$, where ${\bf g}_{ru}\sim\mathcal{CN}({\bf 0},{\bf I}_M)$ accounts for the small-scale fading, and as noted earlier, $\psi_{ru}$ and $\ell(d_{ru})$ account for the shadowing and path loss, respectively. 

We define the $( M|\mathcal{C}_{u'}| \times 1)$ complex vector ${\bf h}_{u',u}$ as the concatenation of the channels $\{ {\bf \bar{h}}_{ru} : r \in {\cal C}_{u'} \}$ between the cluster $\mathcal{C}_{u'}$ serving user $u'$ and the user $u$. In a similar fashion, we define ${\bf w}_{u}$ as the concatenation of the beamformers $\{{\bf \bar{w}}_{ru} : r \in \mathcal{C}_u\}$ used to serve user $u$ by its serving cluster $\mathcal{C}_{u}$. Finally, we use the block diagonal matrix ${\bf S}_{u} = \left({\rm diag} \left( \{ {s}_{ru} \}_{r \in \mathcal{C}_u} \right) \otimes {\bf I}_M \right)$, where ${s}_{ru}$ represents the scheduling of user $u$ by RRH $r \in \mathcal{C}_u$. Specifically, $s_{ru} = 1\ (\mathrm{or~} 0)$ if user $u$ is scheduled (or not scheduled) by RRH $r \in \mathcal{C}_u$. We note that a user $u$ may be scheduled by only some of the RRHs within its serving cluster~$\mathcal{C}_u$.


\subsection{Channel Estimation}
Our optimizaton framework will depend on availability of CSI. To obtain CSI, we consider a time-division duplex (TDD) system with an uplink training phase of length $\tau_p$ used to estimate the channels. During uplink training, the signal ${\bf Y}_r \in \mathbb{C}^{M \times \tau_p}$ received at RRH $r$ is given by
\begin{align}\label{eq:receivedPilotSig}
	{\bf Y}_r = \sum_{u \in \mathcal{U}} \sqrt{p_u} {\bf \bar{h}}_{ru} {\bm \phi}_u + {\bf Z}_r,
\end{align}
where ${\bm \phi}_u \in \mathbb{C}^{1 \times \tau_p}$ is the unit norm (${\bm \phi}_u {\bm \phi}_u^H = 1$) pilot sequence used by user $u$, $p_u$ is the training power used by user $u$, and ${\bf Z}_r$ is the noise with independent and identically distributed entries, distributed as $\mathcal{CN}\left(0, \sigma_Z^2 \right)$. As in~\cite{cellFreeVersusSmallCells7827017, cellFreeUserCentricPower8901451}, we assume the knowledge of the transmit power of the users and the large-scale fading. Hence, for the vector ${\bf \breve{y}}_r = {\rm vec}\left({\bf Y}_r\right) \in \mathbb{C}^{M \tau_p  \times 1}$, the channels $\{{\bf \hat{h}}_{ru} : u \in \mathcal{E}_r\}$ can be estimated using linear minimum MSE (MMSE) as~\cite{kay1993fundamentals}
\begin{align}
	{\bf \hat{h}}_{ru} = {\bf R}_{ru} {\bf R}_{r}^{-1} {\bf \breve{y}}_r,
\end{align}
\vspace{-1em}
\begin{flalign}
	\text{with}\ 
	{\bf R}_{ru} &= \sqrt{p_u} \psi_{ru} \ell(d_{ru}) \left( {\bm \phi}_u^* \otimes {\bf I}_M \right) \\
	\text{and~~}\ 
	{\bf R}_{r} &= \sum_{u \in \mathcal{U}} p_u \psi_{ru} \ell(d_{ru}) \left( {\bm \phi}_u^T {\bm \phi}_u^* \otimes {\bf I}_M \right) + \sigma_z^2 {\bf I}_{M \tau_p}
\end{flalign}

If $\tau_p \ge |\mathcal{U}|$, the users' pilot sequences are orthogonal, and we can avoid inter-user interference in the training phase. Unfortunately, given the large number of users in the network, this is not always feasible, leading to pilot contamination. To address pilot contamination, we group users within $\tau_p$ users per group. Within a group, users use orthogonal training sequences thereby eliminating intra-group contamination, but users in different groups share pilot sequences. We elaborate on how we groups users below.

Using MMSE estimation, the estimated channel satisfies ${\bf \hat{h}}_{ru} \sim \mathcal{CN}\left( {\bf 0}, {\bm \Psi}_{ru} \right)$, where
\begin{align}
	{\bm \Psi}_{ru} = {\bf D}_{ru} \left( \sum_{u' \in \mathcal{U}_u} {\bf D}_{ru'} + \frac{\sigma_Z^2}{p_u} {\bf I}_M \right)^{-1} {\bf D}_{ru},
\end{align}
where ${\bf D}_{ru} \in \mathbb{C}^{M \times M}$ is a diagonal matrix with entries $\left[{\bf D}_{ru}\right]_{mm} \triangleq \psi_{ru} \ell(d_{ru})$, and $\mathcal{U}_u$ is the set of users using the same pilot as user $u$ (including user $u$). As channels are estimated using linear MMSE, the channel estimation error $\mathrm{ {\bf \bar{e}}}_{ru} = {\bf \bar{h}}_{ru} - {\bf \hat{h}}_{ru}$ is uncorrelated with ${\bf \hat{h}}_{ru}$, and it can be modeled as $\mathrm{ {\bf \bar{e}}}_{ru} \sim \mathcal{CN}\left({\bf 0}, {\bm \Theta_{ru}}\right)$, where ${\bm \Theta_{ru}} \triangleq {\bf D}_{ru} - {\bm \Psi}_{ru}$.

Explicitly including channel estimation error, the signal received at user $u$ can be written as
\begin{align}\label{eq:signalModel_imperfectChan}
	y_{u} &=
	\sum_{r \in \mathcal{C}_u} \sqrt{{s}_{ru}}
	\left( {\bf \hat{h}}_{ru}^H + \mathrm{ {\bf \bar{e}}}_{ru}^H \right) {\bf w}_{ru} x_{u}
	\nonumber \\
	&\ 
	+ \sum_{u' \in \mathcal{U}_{-u}} \sum_{r' \in \mathcal{C}_{u'}} \sqrt{{s}_{r'u'}} \left( {\bf \hat{h}}_{r'u}^H + \mathrm{ {\bf \bar{e}}}_{r'u}^H \right) {\bf w}_{r'u'} x_{u'}
	+ z_{u}
\end{align}

Due to the multiple uncertain terms in the received signal, it is difficult to obtain an accurate closed-form expression of the data rate of the users, since we still do not have a closed-form expression for the optimized beamformer. Similar to~\cite{CoherentCRAN8606433}, we consider a lower-bound for the data rate, which allows us to form a tractable expression for the SINR using Jensen's Inequality. 
This lower-bound for the axuliary data rate expression is derived using an expectation over the unknown instantaneous CSI error $\{\mathrm{ {\bf e}}_{u',u}: u, u' \in \mathcal{U}\}$; specifically, $\mathbb{E}_{\mathrm{ {\bf e}}}\left\{\log\left(1 + \gamma_{u} \right) \right\} \ge \log\left(1 + 1/\mathbb{E}_{\mathrm{ {\bf e}}}\left\{\gamma_{u}^{-1} \right\} \right)$. In this approach, we use an auxiliary variable $\gamma_{u}$ for the SINR for the coherent transmission case that is defined as~\cite{CoherentCRAN8606433, 8247283}
\begin{align}
	\gamma_{u}
	&=
	\frac{ 
		{\bf w}_{u}^H {\bf \widetilde{h}}_{u,u} {\bf \widetilde{h}}_{u,u}^H {\bf w}_{u} }
	{ C_{u}
	},
\label{eq:gamma_opt_Coher_est}
\end{align}
with
\begin{align}
	&C_{u}
	=
	\resizebox{0.88\columnwidth}{!}
	{$\displaystyle
		\sum_{u' \in \mathcal{U}_{-u}} {\bf w}_{u'}^H {\bf \widetilde{h}}_{u',u} {\bf \widetilde{h}}_{u',u}^H {\bf w}_{u'}
		+ \sum_{u' \in \mathcal{U}_{-u}} \mathbb{E}\bigg\{ {\bf w}_{u'}^H \mathrm{ {\bf e}}_{u',u} \mathrm{ {\bf e}}_{u',u}^H {\bf w}_{u'} \bigg\}
	$}
	\nonumber\\
	&\ 
	+ \mathbb{E}\bigg\{ {\bf w}_{u}^H \mathrm{ {\bf e}}_{u,u} \mathrm{ {\bf e}}_{u,u}^H {\bf w}_{u} \bigg\}
	+ \sigma_z^2
	\nonumber \\
	&=
	\resizebox{0.85\columnwidth}{!}
	{$\displaystyle
	\sum_{u' \in \mathcal{U}_{-u}} {\bf w}_{u'}^H {\bf \widetilde{h}}_{u',u} {\bf \widetilde{h}}_{u',u}^H {\bf w}_{u'}
	+ \sum_{u' \in \mathcal{U}} {\bf w}_{u'}^H {\bm \Theta}_{u',u} {\bf w}_{u'}
	+ \sigma_z^2
	$}
	,
\end{align}
where the $( M|\mathcal{C}_{u}| \times 1)$ complex vector ${\bf \widetilde{h}}_{u',u}$ is the concatenation of the estimated channels $\left\{ {\bf \widetilde{h}}_{ru} : r \in {\cal C}_{u'} \right\}$ between the cluster $\mathcal{C}_{u'}$ serving user $u'$ and the user $u$. Furthermore, $\mathbb{E}\{ \mathrm{ {\bf e}}_{u',u} \mathrm{ {\bf e}}_{u',u}^H \} = {\bm \Theta_{u',u}}$ is the covariance matrix of the estimation error for the concatenated channels between the RRHs $r \in \mathcal{C}_{u'}$ and user $u \in \mathcal{U}$.
\begin{remark}\label{remark:coherentSINR}
We will be using the auxiliary variable  in~\eqref{eq:gamma_opt_Coher_est} to optimize both the transmit beamforming and the scheduling for the users. This formula assumes that the user has perfect knowledge of the effective precoded channel ${\bf \widetilde{h}}_{u,u}^H {\bf w}_{u}$. For this assumption to hold in practice, some downlink pilot resources are required to train the effective precoded channels, which would incur some estimation error. However as shown in~\cite{5466522}, the effect of this error is marginal compared to the impact of the channel estimation error considered here. More importantly, for our purposes, we would not be able to analyze this error because the beamformers are unknown. In turn, again, because we do not enforce a specific beamforming scheme, we cannot use the capacity bounds used in~\cite{cellFreeUserCentricPower8901451} that take into account the channel uncertainity at the receiver. The difference in performance when the user knows the effective channel or only the channel statistics is quantized in~\cite{cellFreeVersusSmallCells7827017}, where a gap less than $8\%$ in the achievable rate is observed when using conjugate beamforming. We note that the evaluation of the performance includes the (substantially larger) rate penalty for uplink channel estimation; if required, a small additional penalty can be added to account for downlink training.
\end{remark}

\subsection{Pilot Assignment Policy}
It is well accepted that a good pilot assignment (PA) policy is pivotal to control pilot contamination~\cite{randomVsStructuredPilots8403508}. While any reasonable scheme is valid, in this paper, we use a heuristic low-overhead PA policy that assigns the same pilots to users that are as far from each other as possible. This approach is reasonable because it reduces the pilot contamination by exploiting the path loss. Specifically, we assign the non-orthogonal pilots using a variant of the hierarchical agglomerative clustering (HAC) algorithm~\cite{gowda1978agglomerative} which creates a cluster tree or dendrogram. The algorithm, described in Algorithm~\ref{tab:PA}, works for an arbitrary number of users.
\begin{algorithm}
	\caption{Proposed Pilot Assignment in User-centric Cell-free MIMO Network} \label{tab:PA}
	{\small
	Treat each active user as a cluster head. \\
	Combine the two nearest clusters into one using an average linkage, here distance between the center of the clusters. \\
	Repeat Step 2 until reaching the root of the tree where all the users are in the same cluster. \\
	While backtracking the tree starting from the root, define each cluster when its number of users is less than or equal to $\tau_p$. \\
	Assign the orthogonal pilots for the users inside each cluster randomly.\label{step:HAC_assignP}
	}
\end{algorithm}

\begin{remark}
It is worth noting that distance-based PA policies have been used in~\cite{pilotAssign9178782}, where the PA inside each cluster (Step~\ref{step:HAC_assignP} in  Algorithm~\ref{tab:PA}) is optimized using a combinatorial search algorithm that is based on the achievable uplink or downlink SE. Unlike these previous studies, optimizing the PA inside each cluster is highly complex for our scheduling problem because the users to be served are not selected \textit{a priori} and the SE is still not computed. Hence this optimization which would require the inclusion of the PA step \textit{inside} our optimization algorithm. This is impractical because users would need to exchange pilot signals until the optimal PA is reached. Therefore, our proposed approach serves as a lower bound for a possible optimized PA policy; importantly, our proposed approach is easy to implement. For further discussion, readers are encouraged to refer to works that specifically study this topic, e.g.,~\cite{GraphCodePilotContamin8487005, ashikhmin2017pilot, pilotAssign9178782, differentCooperationLevels8845768}.
\end{remark}

\begin{remark}\label{remark:HAC}
The HAC algorithm has a complexity of $\mathcal{O}\left(|\mathcal{U}|^3\right)$ which is higher than algorithms like the K-means and Gaussian mixture models that have linear complexity. However, unlike these algorithms, the HAC is consistent and is robust to the choice of distance metric~\cite{karypis2000comparison}. Furthermore, it does not require the prior selection of the number of clusters, allowing us to define the clusters based on an upper limit on the number of users allowed per cluster (the pilot sequence length, $\tau_p$). Additionally, solutions like~\cite{randomVsStructuredPilots8403508}, which use K-means clustering to determine the co-pilot users (through the centroids of the clusters), do not provide a robust method to obtain the required number of clusters, especially when the number of users is not a multiple of the length of the pilot sequence. This causes a problem, because unless we implement a procedure to dynamically change the number of clusters each time we assign the co-pilot users, we will not be able to assign pilot sequences to all the users while having comparable number of users for each pilot sequence, or we may end up having some pilot sequences wasted and not assigned to any user.
\end{remark}

\section{Coherent Transmission}\label{section:Formulation}
\subsection{Problem Definition}\label{section:prob_definition}
Given the network and signal models, we are now ready to formulate the weighted sum-rate optimization problem over the beamformers and scheduling variables. The problem at hand is 
\begin{subequations}\label{eq:TotalProblem_1}
	\begin{align}
	&\stepcounter{probNum}
	(\mathrm{P}\arabic{probNum})\quad
	\max_{ {\mathcal{S}}, {\mathcal{W}}}\quad \sum_{u \in \mathcal{U}} \delta_{u}
	\log\left( 1 + \gamma_{u} \right) 
	\label{eq:TotalProblem_1_obj}
	\\
	&\text{s.t.}\quad 
	\sum_{u \in \mathcal{E}_r} {s}_{ru} \le M,
	\mkern200mu
	r \in \mathcal{B}
	\label{eq:TotalUtilityProblem_U_1}
	\\
	&\quad\quad
	\sum_{u\in \mathcal{E}_r} \|{\bf \bar{w}}_{ru}\|_2^2 \le p,
	\mkern180mu
	r \in \mathcal{B} 
	\label{eq:TotalUtilityProblem_U_1_w}
	\\
	&\gamma_{u} = 
	\frac{ 
		\resizebox{0.37\columnwidth}{!}
		{$
		\displaystyle
		{\bf w}_{u}^H {\bf S}_{u}^{1/2} {\bf \widetilde{h}}_{u,u} {\bf \widetilde{h}}_{u,u}^H {\bf S}_{u}^{1/2} {\bf w}_{u} 
		$}
	}
	{ 
		\resizebox{0.87\columnwidth}{!}
		{$
		\displaystyle
		\sum_{u' \in \mathcal{U}_{-u}} {\bf w}_{u'}^H {\bf S}_{u'}^{1/2} {\bf \widetilde{h}}_{u',u} {\bf \widetilde{h}}_{u',u}^H {\bf S}_{u'}^{1/2} {\bf w}_{u'} 
		+ \sum_{u' \in \mathcal{U}} {\bf w}_{u'}^H {\bm \Theta}_{u',u} {\bf w}_{u'}
		+ \sigma_z^2
		$}
	}
	,
	\nonumber \\
	&
	\mkern350mu
	u \in \mathcal{U} 
	\label{eq:TotalUtilityProblem_U_1_SINR}
	\\
	&\quad\quad
	{s}_{ru} \in \{0, 1\},
	\mkern165mu
	r \in \mathcal{B}, u \in \mathcal{U}
	\label{eq:TotalUtilityProblem_U_1_tau}
	\end{align}
\end{subequations}
The variables $\mathcal{S} =\{ {\bf S}_1, \dots, {\bf S}_{|\mathcal{U}|} \}$ denote the scheduling variables defined earlier for the users. Similarly, $\mathcal{W} =  \{{\bf w}_{1},\ \dots,\ {\bf w}_{|\mathcal{U}|} \}$ is the set of the beamformers constructed to serve the users. $\delta_{u}$ represents the weight assigned to user $u$, used to provide fairness. In~\eqref{eq:TotalProblem_1} we wish to optimize the scheduling variables ${\mathcal{S}}$ and beamformers ${\mathcal{W}}$ such that the total network utility in~\eqref{eq:TotalProblem_1_obj}, the WSR, is maximized.  Constraint~\eqref{eq:TotalUtilityProblem_U_1} specifies that an RRH with $M$ antennas serves at most $M$ users, even though it may be associated with more than $M$ users ($|\mathcal{E}_r| > M$). The beamformers implicitly include power allocation and constraint~\eqref{eq:TotalUtilityProblem_U_1_w} specifies the power budget of the RRHs. Constraint~\eqref{eq:TotalUtilityProblem_U_1_tau} shows that user $u$ may either be scheduled or not. Constraint~\eqref{eq:TotalUtilityProblem_U_1_SINR} defines the SINR for the users. As suggested in Remark~\ref{remark:coherentSINR}, we use a proxy for the SINR and, hence, rate.

Problem~\eqref{eq:TotalProblem_1} is a mixed-integer non-convex problem due to the included binary variables and their presence in both the numerators and the denominators in the utility function. As such, it is hard to solve in its current form; similar WSR resource allocation problems have been proven to be NP-hard~\cite{Complexity4453890}. Hence, obtaining a global optimum is in most cases computationally prohibitive.

\subsection{Problem Analysis for Coherent Transmission}\label{sectionc:Analysis}
Here, we analyze the problem at hand and set up the development of the algorithm in the next section. The user scheduling component of problem~\eqref{eq:TotalProblem_1} warrants further discussion. At each RRH, we have a maximum of $M$ non-zero beamformers. Given the beamformers, to optimize the scheduling parameters in such a case, we need to solve a combinatorial problem whereby the available non-zero beams are assigned to the users such that the network utility is maximized. This is a problem of matching beamformers to users, which, for the case of a co-located transmitter, can be expressed as finding a permutation matrix $\mathbf{P}$ to maximize $\mathtt{tr}\{\mathbf{A}\mathbf{P}\}$. Here, $\mathbf{A}$ would be the matrix of utilities, i.e., its entry $A_{ub}$ would the utility of user $u$ if beam $b$ is assigned to user $u$ by its single serving transmitter. For co-located transmitters, this problem can be efficiently solved using, e.g., the Hungarian algorithm~\cite{Ahmad9084256} with polynomial complexity.

Unfortunately, when the user can be served by many RRHs, the problem is coupled across RRHs and the matching problem is not a conventional assignment problem~\cite{bokhari2012assignment}. Specifically, unlike in~\cite{Ahmad9084256}, if a user's beam assignment at one RRH changes, the interference experienced by this user changes. Consequently, the weighted rate (utility) of each user is based on the \textit{combination} of beam assignment decisions at multiple RRHs. We are therefore forced to explore alternative solutions; in this regard, we use ideas from compressive sensing.

We begin the analysis by rewriting the constraint in~\eqref{eq:TotalUtilityProblem_U_1}. Using the fact that the beamforming vectors are constructed only for users actually served, we have $
{s}_{ru} \triangleq \mathbbm{1} \{ \|{\bf \bar{w}}_{ru}\|_2^2 \}$ where $\mathbbm{1} \{\cdot\}$ is the indicator function ($\mathbbm{1}(x) = 1, x > 0$, else $= 0$). 

The problem in~\eqref{eq:TotalProblem_1} can now be rewritten as
\begin{subequations}\label{eq:TotalUtilityProblem_w}
	\begin{align}
	&\stepcounter{probNum}
	(\mathrm{P}\arabic{probNum})\quad
	\max_{ \mathcal{W}}\quad \sum_{u \in \mathcal{U}}  	
	\delta_{u}
	\log\left( 1 + 
	\gamma_{u}
	\right)
	\label{eq:TotalUtilityProblem_w_obj}
	\\
	&\text{s.t.}\quad 
	\sum_{u \in \mathcal{E}_r} \mathbbm{1} \{ \|{\bf \bar{w}}_{ru}\|_2^2 \} \le M,
	\mkern130mu
	r \in \mathcal{B}
	\label{eq:TotalUtilityProblem_w_indc1}
	\\
	&\quad \quad
	\sum_{u\in \mathcal{E}_r} \|{\bf \bar{w}}_{ru}\|_2^2 \le p,
	\mkern170mu
	r \in \mathcal{B}
	\label{eq:TotalUtilityProblem_w_powerBudget}
	\\
	&
	\gamma_{u} = 
	\frac{ 
		\displaystyle
		{\bf w}_{u}^H {\bf \widetilde{h}}_{u,u} {\bf \widetilde{h}}_{u,u}^H {\bf w}_{u} 
	}
	{ 
		\displaystyle
		\sum_{u' \in \mathcal{U}_{-u}} {\bf w}_{u'}^H {\bf \widetilde{h}}_{u',u} {\bf \widetilde{h}}_{u',u}^H {\bf w}_{u'} 
		+ \sum_{u' \in \mathcal{U}} {\bf w}_{u'}^H {\bm \Theta}_{u',u} {\bf w}_{u'}
		+ \sigma_z^2
	},
	\nonumber \\
	&
	\mkern340mu
	u \in \mathcal{U}
	\label{eq:TotalUtilityProblem_w_SINR}
	\end{align}
\end{subequations}
where the optimization variable is, now, the set $\mathcal{W}$ which represents the beamformers for the RRHs. The constraint in~\eqref{eq:TotalUtilityProblem_w_SINR} does not include the indicator function of the beamformer, because it already includes the beamformers $\mathcal{W}$, which will be zero for the unscheduled users.

Using the objective function in~\eqref{eq:TotalUtilityProblem_w_obj} and the equality constraints in \eqref{eq:TotalUtilityProblem_w_SINR}, we can write the following Lagrangian formulation
\begin{align}\label{eq:Lagrangian}
&\mathcal{L}(\mathcal{W}, {\bm \gamma}, {\bm \nu}) = 
\sum_{u \in \mathcal{U}} \delta_{u}
\log\left( 1 + 
\gamma_{u}
\right) - \sum_{u \in \mathcal{U}} \nu_u \Bigg( \gamma_{u}
\nonumber \\
&
- \frac{ 
	{\bf w}_{u}^H {\bf \widetilde{h}}_{u,u} {\bf \widetilde{h}}_{u,u}^H {\bf w}_{u} }
{ \displaystyle
	\sum_{u' \in \mathcal{U}_{-u}} {\bf w}_{u'}^H {\bf \widetilde{h}}_{u',u} {\bf \widetilde{h}}_{u',u}^H {\bf w}_{u'} 
	+ \sum_{u' \in \mathcal{U}} {\bf w}_{u'}^H {\bm \Theta}_{u',u} {\bf w}_{u'} + \sigma_z^2} \Bigg),
\end{align}
where the vector ${\bm \gamma} = [\gamma_{1}\ \dots\ \gamma_{|\mathcal{U}|}]^T$ represents the SINR auxiliary variables and ${\bm \nu}$ denotes the associated Lagrange multipliers. By fixing $\mathcal{W}$, we can obtain an expression for the Lagrange multiplier by setting the derivative of~\eqref{eq:Lagrangian} with respect to $\gamma_{u}$ to zero. This results in
\begin{equation}\label{eq:LagrangeSINR}
\nu_u =
\resizebox{0.9\columnwidth}{!}
{$ \displaystyle 
\frac{ \displaystyle
		\delta_{u} \bigg( \sum_{u' \in \mathcal{U}_{-u}} {\bf w}_{u'}^H {\bf \widetilde{h}}_{u',u} {\bf \widetilde{h}}_{u',u}^H {\bf w}_{u'} 
		+ \sum_{u' \in \mathcal{U}} {\bf w}_{u'}^H {\bm \Theta}_{u',u} {\bf w}_{u'} + \sigma_z^2 \bigg)
	}
	{ \displaystyle
		\sum_{u' \in \mathcal{U}} {\bf w}_{u'}^H  \left( {\bf \widetilde{h}}_{u',u} {\bf \widetilde{h}}_{u',u}^H + {\bm \Theta}_{u',u} \right) {\bf w}_{u'} 
		+ \sigma_z^2
	}
$}
\end{equation}
Using~\eqref{eq:LagrangeSINR} in~\eqref{eq:Lagrangian}, we can reformulate our objective function in~\eqref{eq:TotalUtilityProblem_w}~as
\begin{align}\label{eq:objectiveFLag}
&f_1(\mathcal{W}, {\bm \gamma})
=
\sum_{u \in \mathcal{U}} \delta_{u}
\left(
\log\left( 1 + 
\gamma_{u}
\right)
- \gamma_{u}
\right)
\nonumber \\
&
+
\resizebox{0.83\columnwidth}{!}
{$ \displaystyle
\sum_{u \in \mathcal{U}}
\delta_{u}
\Bigg( \frac{ \left(1 + \gamma_{u}\right) {\bf w}_{u}^H {\bf \widetilde{h}}_{u,u} {\bf \widetilde{h}}_{u,u}^H {\bf w}_{u} }
{ \displaystyle
	\sum_{u' \in \mathcal{U}} {\bf w}_{u'}^H  \left( {\bf \widetilde{h}}_{u',u} {\bf \widetilde{h}}_{u',u}^H + {\bm \Theta}_{u',u} \right) {\bf w}_{u'} + \sigma_z^2}
\Bigg)
$}
\end{align}
As expected, setting the derivative of~\eqref{eq:objectiveFLag} to zero, we re-create the equality constraint in~\eqref{eq:TotalUtilityProblem_w_SINR}.
If we substitute this optimal value of $\gamma_{u}$ into~\eqref{eq:objectiveFLag} we obtain the same objective function in~\eqref{eq:TotalUtilityProblem_w}, i.e., the function in~\eqref{eq:objectiveFLag} is equivalent to the objective function in~\eqref{eq:TotalUtilityProblem_w}. Hence, the optimization in~\eqref{eq:TotalUtilityProblem_w} can be written~as
	\begin{align}\label{eq:TotalProblem_lag_problem}
	\stepcounter{probNum}
	(\mathrm{P}\arabic{probNum})\quad
	\max_{ \mathcal{W}, {\bm \gamma}}\quad & 
	f_1( \mathcal{W}, {\bm \gamma})
\quad\quad
	\text{s.t.}\quad  
	\eqref{eq:TotalUtilityProblem_w_indc1}, \eqref{eq:TotalUtilityProblem_w_powerBudget}
	\end{align}
The importance of this procedure is that it allows us to write the optimization terms other than ${\bm \gamma}$ outside the logarithm function simplifying the formulation of an algorithm. We emphasize that the $(\mathrm{P}\arabic{probNum})$ is not the dual problem of~\eqref{eq:TotalUtilityProblem_w}, but rather we have introduced SINR auxiliary variables ${\bm \gamma}$ that act as a proxy to simplify algorithm design. 

The optimization variables found in the objective function of problem~\eqref{eq:TotalProblem_lag_problem} are found in both the numerator and the denominator, which makes the problem hard to solve. 
Applying the fractional programming approach~\cite[Corollary~1]{FR8310563} on the objective function~\eqref{eq:objectiveFLag}, we can write the following function
\begin{align}\label{eq:objectiveFLag_c_term_linearized}
	&f_2(\mathcal{W}, {\bm \gamma}, {\bm \beta})
	=
	\sum_{u \in \mathcal{U}} \delta_{u}
	\left(
	\log\left( 1 + 
	\gamma_{u}
	\right)
	- \gamma_{u}
	\right)
	\nonumber \\[-5pt]
	&
	+
	\sum_{u \in \mathcal{U}}
	\Bigg(
	2 \text{Re}\left\{
	\beta_{u}^{*}
	\sqrt{\delta_{u} \left(1 + \gamma_{u}\right) }
	{\bf w}_{u}^H {\bf \widetilde{h}}_{u,u}
	\right\}
	\nonumber \\[-5pt]
	&
	-
	|\beta_{u}|^2
	\left(
	\sum_{u' \in \mathcal{U}} {\bf w}_{u'}^H  \left( {\bf \widetilde{h}}_{u',u} {\bf \widetilde{h}}_{u',u}^H + {\bm \Theta}_{u',u} \right) {\bf w}_{u'} + \sigma_z^2
	\right)
	\Bigg),
\end{align}
where vector ${\bm \beta} = \left[ \beta_{1} \dots \beta_{|\mathcal{U}|} \right]^T$ is introduced as a new auxiliary variable as required by fractional programming, and $\text{Re}\{\cdot\}$ denotes the real part of a complex number. Equation~\eqref{eq:objectiveFLag_c_term_linearized} has a summation form making it easier to solve than that in~\eqref{eq:objectiveFLag}. We note that~\eqref{eq:objectiveFLag_c_term_linearized} is concave in ${\bm \beta}$.
\begin{remark}
	Maximizing the function in~\eqref{eq:objectiveFLag_c_term_linearized} over the new auxiliary variables ${\bm \beta}$ is equivalent to~\eqref{eq:objectiveFLag}, i.e., $f_1( \mathcal{W}, {\bm \gamma}) = \underset{\bm \beta}{\max}\ f_2(\mathcal{W}, {\bm \gamma}, {\bm \beta})$. This can be proved in the same way as was done with~\eqref{eq:objectiveFLag}, i.e., by setting the partial derivative of~\eqref{eq:objectiveFLag_c_term_linearized} with respect to $\beta_{u}$ to zero and then substituting the resulting value of $\beta_{u}$ into~\eqref{eq:objectiveFLag_c_term_linearized}, which yields~\eqref{eq:objectiveFLag}.
\end{remark}

\section{Resource Allocation Algorithm for Coherent Transmission}\label{section:ResourceAllocation}
In this section, we develop our algorithm to address the user scheduling and resource allocation problem assuming ideal CSI at the RRHs. We note that, since the $\mathrm{l}_0$-norm of a vector counts its number of non-zero elements, the indicator function found in~\eqref{eq:TotalUtilityProblem_w_indc1} can be further replaced by an $\mathrm{l}_0$-norm of the $\left\|{\bf \bar{w}}_{ru} \right\|_2^2$, i.e., $\mathbbm{1} \left\{ \left\|{\bf \bar{w}}_{ru} \right\|_2^2 \right\} = \left\| \left\| {\bf \bar{w}}_{ru} \right\|_2^2 \right\|_0$. Unfortunately, this change, in of itself, does not provide a mathematically tractable problem. However, the $\mathrm{l}_0$-norm of a vector ${\bf x}$ can be approximated as a weighted convex $\mathrm{l}_1$-norm as~\cite{candes2008enhancing}
\begin{align}\label{eq:L0}
\|{\bf x}\|_0 \simeq \sum_{m} \alpha_m |x_m| = \| {\bm \alpha} {\bf x}\|_1,
\end{align}
where $\alpha_m$ are positive weights that penalize the nonzero coefficients $x_m$, and ${\bm \alpha} = {\rm diag}\{\alpha_1, \alpha_2, \dots\}$ is a diagonal matrix. For our case, ${\bf x} = \left\| {\bf \bar{w}}_{ru} \right\|_2$ which is scalar, hence we have one weight, $\alpha_{ru}$, per constraint in~\eqref{eq:TotalUtilityProblem_w_indc1}. Hence for our network, we can construct an iterative process to find these weights at each iteration $j$ using~\cite{candes2008enhancing}
\begin{align}\label{eq:weightsUpdate}
\alpha_{ru}^{(j+1)} = \frac{1}{ \big\| {\bf \bar{w}}_{ru}^{(j)} \big\|_2^2 + \epsilon} \ ,
\end{align}
where ${\bf \bar{w}}_{ru}^{(j)}$ is the value of ${\bf \bar{w}}_{ru}$ from current iteration, and it is used to update $\alpha_{ru}^{(j+1)}$, i.e., the value of $\alpha_{ru}$ in the next iteration. The term $\epsilon > 0$ provides stability.
At its core, a large value of $\alpha_{ru}$ encourages $\left\| {\bf \bar{w}}_{ru} \right\|_2^2$ toward zero. Generally, $\epsilon$ is chosen to be slightly smaller than the expected value of $\left\| {\bf \bar{w}}_{ru} \right\|_2^2$ for the scheduled users~\cite{candes2008enhancing}. In our testing, we found that the performance of our algorithm was not particularly sensitive to reasonable choices of $\epsilon$. 

The approximation in~\eqref{eq:L0} allows us to rewrite constraint~\eqref{eq:TotalUtilityProblem_w_indc1} as a convex function, which, in turn, allows us to formulate convex problems in each individual variable (the overall problem is \emph{not jointly convex}), i.e., we can use coordinate descent. We now have
\vspace{-0.5em}
\begin{subequations}\label{eq:TotalUtilityProblem_w_weights}
	\begin{align}
	\stepcounter{probNum}
	(\mathrm{P}\arabic{probNum})\quad
	\max_{ \mathcal{W}, {\bm \gamma}, {\bm \beta}}\quad & f_2( \mathcal{W}, {\bm \gamma}, {\bm \beta})
	&
	\label{eq:TotalUtilityProblem_w_weights_obj}
	\\
	\text{s.t.}\quad 
	& 
	\sum_{u \in \mathcal{E}_r} \alpha_{ru} \|{\bf \bar{w}}_{ru}\|_2^2 \le M,
	\label{eq:TotalUtilityProblem_w_weights_M}
	&
	r \in \mathcal{B}
	\\
	&
	\sum_{u\in \mathcal{E}_r} \|{\bf \bar{w}}_{ru}\|_2^2 \le p,
	\label{eq:TotalUtilityProblem_w_powerB}
	&
	r \in \mathcal{B}
	\end{align}
\end{subequations}
\vspace{-1.8em}
\begin{remark}
when ${\bm \gamma}$ and ${\bm \beta}$ are fixed, \eqref{eq:TotalUtilityProblem_w_weights}~is quadratically constrained quadratic programming (QCQP) problem in $\mathcal{W}$; the capacity constraint in~\eqref{eq:TotalUtilityProblem_w_weights_M} is also a weighted power constraint.
\end{remark}

\vspace{-2em}
\subsection{Optimal Expressions for Individual Variables}
When the variables $\mathcal{W}$ and ${\bm \gamma}$ are fixed, the optimal value of the auxiliary variable $\beta_{u}$ can be obtained from its corresponding first optimality condition from~\eqref{eq:objectiveFLag_c_term_linearized}~as
\begin{align}\label{eq:beta_opt_Coher}
\beta_{u}
=
\frac{
	\sqrt{\delta_{u}\left( 1 + \gamma_{u}\right)}
	{\bf w}_{u}^H {\bf \widetilde{h}}_{u,u}
}
{
	\sum_{u' \in \mathcal{U}} {\bf w}_{u'}^H  \left( {\bf \widetilde{h}}_{u',u} {\bf \widetilde{h}}_{u',u}^H + {\bm \Theta}_{u',u} \right) {\bf w}_{u'} + \sigma_z^2
}
\end{align}
To obtain a similar expression for the beamformer, we first define the following function
\vspace{-0.5em}
\begin{align}
&f_3(\mathcal{W}, {\bm \mu}, {\bm \lambda})
=
-
\sum_{r \in \mathcal{B}} \mu_r \bigg( 
\sum_{u \in \mathcal{E}_r} \|{\bf \bar{w}}_{ru}\|_2^2 - p
\bigg)
\nonumber\\
&\quad\quad\quad
-
\sum_{r \in \mathcal{B}} 
\lambda_r
\bigg(
\sum_{u \in \mathcal{E}_r} 
\alpha_{ru} \|{\bf \bar{w}}_{ru}\|_2^2
- M
\bigg)
\nonumber \\
&
=
\resizebox{0.83\columnwidth}{!}
{$\displaystyle
-
\sum_{u \in \mathcal{U}}
\sum_{r \in \mathcal{C}_u}
\left(\mu_r + \lambda_r \alpha_{ru}\right)\|{\bf \bar{w}}_{ru}\|_2^2
+ \sum_{r \in \mathcal{B}} \left( p \mu_r + M \lambda_r\right)
$}
\end{align}
where $\mu_{ru}$ and $\lambda_{ru}$ are the Lagrange multipliers corresponding to the constraints in~\eqref{eq:TotalUtilityProblem_w_weights_M} and~\eqref{eq:TotalUtilityProblem_w_powerB} respectively. We can then write the Lagrangian formulation of~\eqref{eq:TotalUtilityProblem_w_weights} as
\vspace{-0.5em}
\begin{align}\label{eq:LagrangianJTDPS_weights_Coher}
&\mathcal{L}(\mathcal{W}, 
{\bm \mu}, {\bm \lambda})
=
\sum_{u \in \mathcal{U}} \delta_{u}
\left(
\log\left( 1 + 
\gamma_{u}
\right)
- \gamma_{u}
\right)
\nonumber \\
&\ 
+ 
\sum_{u \in \mathcal{U}}
\Bigg(
2 \text{Re}\left\{
\beta_{u}^{*}
\sqrt{\delta_{u} \left(1 + \gamma_{u}\right) }
{\bf w}_{u}^H {\bf \widetilde{h}}_{u,u}
\right\}
\nonumber \\
&\ 
-
|\beta_{u}|^2
\left(
\sum_{u' \in \mathcal{U}} {\bf w}_{u'}^H  \left( {\bf \widetilde{h}}_{u',u} {\bf \widetilde{h}}_{u',u}^H + {\bm \Theta}_{u',u} \right) {\bf w}_{u'} + \sigma_z^2
\right)
\nonumber \\
&\ 
+ f_3(\mathcal{W}, {\bm \mu}, {\bm \lambda})
\end{align}

Considering the first optimality condition for~\eqref{eq:LagrangianJTDPS_weights_Coher}, when both the variables ${\bm \gamma}$ and ${\bm \beta}$ are fixed, the optimal beamformers expression for ${\bf w}_{u}$ can be constructed as
\vspace{-0.5em}
\begin{align}\label{eq:beamformer_opt_Concat_Coher}
&{\bf w}_{u}
=
\sqrt{\delta_{u}\left( 1 + \widetilde{\gamma}_{u}\right)}
\beta_{u}^{*}
\nonumber \\
&
\times
\left(
\sum_{u' \in \mathcal{U}}
|\beta_{u'}|^2
\left( {\bf \widetilde{h}}_{u,u'} {\bf \widetilde{h}}_{u,u'}^H + {\bm \Theta}_{u,u'} \right)
+ {\bf B}_{u} \right)^{-1}
{\bf \widetilde{h}}_{u,u},
\end{align}
where ${\bf B}_{u} = \left({\rm diag} \left( \{ \left(\mu_r + \lambda_r \alpha_{ru}\right) \}_{r \in \mathcal{C}_u} \right) \otimes {\bf I}_M \right)$. This is similar to an MMSE beamformer.

The Lagrangian multipliers $\mu_r \in \mathbb{R}^{+}$ and $\lambda_r \in \mathbb{R}^{+}$, found in ${\bf B}_{u}$, can be determined through the capacity,~\eqref{eq:TotalUtilityProblem_w_weights_M}, and power budget,~\eqref{eq:TotalUtilityProblem_w_powerB}, constraints respectively. Importantly, both these constraints relate to the power used at RRH $r$, i.e., with probability one, both cannot be tight simultaneously. From complementary slackness, therefore, one of these Lagrange	multipliers, both corresponding to RRH $r$, must be zero. 

Unfortunately, we do not know a priori which constraint	will remain tight. As we will see in our algorithm section, we propose a heuristic that, at each iteration of the algorithm, checks for whether the capacity constraint is satisfied (allowing $\lambda_r = 0$); if it is not satisfied, we set $\lambda_r$ to a small value and update $\mu_r$ using a bisection search to meet the power constraint. In our testing, after a few iterations $\lambda_r$ always converges to zero.

\vspace{-0.5em}
\begin{remark}
We will use~\eqref{eq:beamformer_opt_Concat_Coher} to obtain the beam vector for user $u$; importantly, this vector covers all RRHs serving user $u$. On the other hand, $\mu_r$ is obtained at individual RRHs and changing its value would change the beams at other RRHs significantly complicating the process of obtaining the required Lagrange multipliers. However, changing $\mu_r$ through a bisection search does not significantly affect the beams at the other RRHs serving $u$, i.e., $\{ {\bf \bar{w}}_{r'u}: r' \ne r \}$. This is because the off-diagonal entries of the term $\sum_{u' \in \mathcal{U}} |\beta_{u'}|^2 {\bf h}_{u,u'} {\bf h}_{u,u'}^H$ (found in~\eqref{eq:beamformer_opt_Concat_Coher}) are very small compared to the diagonal entries. This claim directly follows from summing up a large number of independent zero-mean random variables.
\end{remark}
\vspace*{-2em}
\subsection{Optimization Algorithm}
We are now ready to state our algorithm. The steps in the algorithm are listed in Algorithm~\ref{algortihm:w_using_weights}. 

\begin{algorithm}
	\SetAlgoLined
	\SetInd{0.1em}{1em}
	\caption{User scheduling and resource allocation}
	\label{algortihm:w_using_weights}
	\small
	Input parameters: $\mathcal{B}$, $ \mathcal{U}$, $\mathcal{E}_r$ and $\mathcal{C}_u$\\
	Initialize $\mathcal{W}$ using conjugate beamforming for \emph{all} users, i.e, ${\bf \bar{w}}_{ru} = \sqrt{\frac{p}{|\mathcal{E}_r|}} \frac{{\bf \bar{h}}_{ru}}{\|{\bf \bar{h}}_{ru}\|}$. \label{step:L_1_norm_weights_init_beams}\\
	Start with weights $\alpha_{ru} = \frac{M}{p}$, $\forall r, u \in \mathcal{E}_r$. \label{step:L_1_norm_weights_init_2}\\
	\While{ \textbf{NOT} converged}{\label{step:Algo_terminate_2}
		Update ${\bm \gamma}$ using~\eqref{eq:TotalUtilityProblem_w_SINR}. \label{step:using_weights_gamma_beta_lambda}
		\\
		Update ${\bm \beta}$ using~\eqref{eq:beta_opt_Coher}.
		\\
		Update $\mathcal{W}$ using~\eqref{eq:beamformer_opt_Concat_Coher}. \label{step:BF} \\
		Update $\{\lambda_r, \mu_r : r \in \mathcal{B}\}$ as described using complementary slackness.\\
		Update weights ${\bm \alpha}$ using~\eqref{eq:weightsUpdate}.\label{step:using_weights_alpha}\\
	}
\end{algorithm}
The algorithm starts by initializing the required variables (Steps~\ref{step:L_1_norm_weights_init_beams}-\ref{step:L_1_norm_weights_init_2}), where we begin with the greedy case of conjugate beamforming for \emph{all} the users in the serving cluster $\mathcal{E}_r$ of each RRH $r$. The algorithm then updates the variables ${\bm \gamma}$, ${\bm \beta}$,  $\mathcal{W}$ and ${\bm \alpha}$ iteratively until convergence. 

\begin{figure}[t]
	\centering
	\vspace{-0.5em}
	\includegraphics[width=0.9\linewidth]{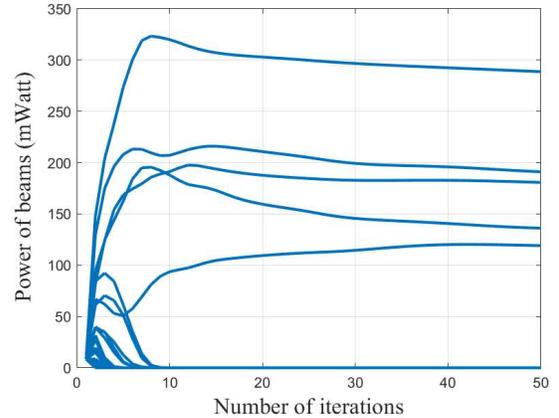}
	\vspace{-0.5em}
	\caption{Evolution of the allocated power for the users' beamformers on a typical RRH.}
	\label{fig:avgSpecEff1}
	\vspace{-1.5em}
\end{figure}

As discussed for $\lambda_r$, at each algorithm iteration we set $\lambda_r$ to either zero or a small constant depending on the capacity constraint being satisfied or not, and we update $\alpha_{ru}$ using~\eqref{eq:weightsUpdate}. Note that, increasing the value of $\lambda_r$ moves the algorithm in the direction of dropping scheduled users because it increases the contribution of ${\bm \alpha}$ in the beamformer expression in~\eqref{eq:beamformer_opt_Concat_Coher}. On the other hand, when $\lambda_r = 0$, the constraint is not active at a specific algorithm iteration. In Fig.~\ref{fig:avgSpecEff1}, we plot the evolution of the allocated power of the beamformer's weights as a function of the algorithm iterations for a typical RRH, where $M = 8$. It is clear that after only a few iterations, the capacity constraint is satisfied.

We note that because at each step (specifically, Steps 5-8), the coordinate descent algorithm provides the optimal solution for one variable holding the others fixed,  the algorithm is guaranteed to be non-decreasing across iterations; we illustrate this convergence in the results section.

\subsection{Resource Allocation Under Perfect CSI}
Based on our derivations in Section~\ref{section:ResourceAllocation}, as in~\eqref{eq:beamformer_opt_Concat_Coher}, under perfect CSI the beamformer can be defined~as
\begin{align}\label{eq:beamformer_opt_ideal}
&{\bf w}_{u, {\rm ideal}}
=
\sqrt{\delta_{u}\left( 1 + \gamma_{u}\right)}
\beta_{u, {\rm ideal}}^{*}
\nonumber \\
&\quad\quad\quad
\times
\left( \sum_{u' \in \mathcal{U}}
|\beta_{u', {\rm ideal}}|^2
{\bf h}_{u,u'} {\bf h}_{u,u'}^H
+ {\bf B}_{u} \right)^{-1}
{\bf h}_{u,u}
\end{align}
Similarly, $\gamma_{u, {\rm ideal}}$ and $\beta_{u, {\rm ideal}}$ are the same as in~\eqref{eq:gamma_opt_Coher_est} and~\eqref{eq:beta_opt_Coher}, respectively, except that we are now using the actual channels instead of the estimated ones, i.e.,
\begin{align}\label{eq:gamma_opt_Coher_ideal}
\gamma_{u, {\rm ideal}}
&=
\frac{ 
	{\bf w}_{u, {\rm ideal}}^H {\bf h}_{u,u} {\bf h}_{u,u}^H {\bf w}_{u, {\rm ideal}} }
{ \displaystyle
	\sum_{u' \in \mathcal{U}_{-u}} {\bf w}_{u', {\rm ideal}}^H {\bf h}_{u',u} {\bf h}_{u',u}^H {\bf w}_{u', {\rm ideal}}
	+ \sigma_z^2
}
&&
\end{align}
\vspace{-1em}
\begin{align}\label{eq:beta_opt_Coher_ideal}
\beta_{u, {\rm ideal}}
=
\frac{
	\sqrt{\delta_{u}\left( 1 + \gamma_{u, {\rm ideal}}\right)}
	{\bf w}_{u, {\rm ideal}}^H  {\bf h}_{u,u}
}
{ \displaystyle
	\sum_{u' \in \mathcal{U}} {\bf w}_{u', {\rm ideal}}^H {\bf h}_{u',u} {\bf h}_{u',u}^H {\bf w}_{u', {\rm ideal}}
	+ \sigma_z^2
}
&&
\end{align}

\begin{table*}[t]
	\begin{align}\label{eq:rate_nonCoher}
		&R_u
		=
		\frac{\left(\tau_d - \tau_p\right)}{\tau_d}
		\log\Bigg(1
		+
		\frac{ 
			\sum_{r \in \mathcal{C}_u} {s}_{ru}
			\left|{\bf \hat{h}}_{ru}^H {\bf \bar{w}}_{ru} \right|^2 }
		{ \displaystyle
			\sum_{r' \in \mathcal{B}} \sum_{u' \in \mathcal{E}_{r'},u' \ne u} {s}_{r'u'} \left| {\bf \hat{h}}_{r'u}^H {\bf \bar{w}}_{r'u'} \right|^2
			+ \sum_{r' \in \mathcal{B}} \sum_{u' \in \mathcal{E}_{r'},u' \ne u} {s}_{r'u'}  {\bf \bar{w}}_{r'u'}^H \bar{\bm \Theta}_{r'u} {\bf \bar{w}}_{r'u'}
			+ \sigma_z^2}
		\Bigg)
		\\ 
		\hline
		\nonumber
	\end{align}
	\vspace{-2.5em}
	\begin{subequations}\label{eq:TotalProblem_decoupledRB}
		\begin{align}
			\stepcounter{probNum}
			(\mathrm{P}\arabic{probNum})\quad
			&\max_{ {\bm {s}}, \mathcal{W}}\quad \sum_{u \in \mathcal{U}} \delta_{u}
			\log\left( 1 + 
			\gamma_{u}
			\right) 
			\\
			\text{s.t.}\quad &
			\gamma_{u} =
			\frac{ \displaystyle
				\sum_{r \in \mathcal{C}_u} {s}_{ru}
				\left|{\bf \hat{h}}_{ru}^H {\bf \bar{w}}_{ru} \right|^2 }
			{ \displaystyle
				\sum_{r' \in \mathcal{B}} \sum_{u' \in \mathcal{E}_{r'},u' \ne u} {s}_{r'u'} \left| {\bf \hat{h}}_{r'u}^H {\bf \bar{w}}_{r'u'} \right|^2
				+ \sum_{r' \in \mathcal{B}} \sum_{u' \in \mathcal{E}_{r'}} {s}_{r'u'}  {\bf \bar{w}}_{r'u'}^H \bar{\bm \Theta}_{r'u} {\bf \bar{w}}_{r'u'}
				+ \sigma_z^2}
			,
			\quad\quad
			u \in \mathcal{U}
			\label{eq:TotalProblem_decoupledRB_SINR}
			\\[5pt]
			&
			\eqref{eq:TotalUtilityProblem_U_1}, \eqref{eq:TotalUtilityProblem_U_1_w}, \eqref{eq:TotalUtilityProblem_U_1_tau} 
			\nonumber
			\\ 
			\hline
			\nonumber
		\end{align}
	\end{subequations}
	\vspace{-3em}
\end{table*}
%
%
\section{Non-coherent Transmission}\label{sec:NonCoher}
For the non-coherent transmission mode, the RRHs transmit different data streams to the same user, and phase synchronization is not required. The data rate of each user is the summation of individual data rates from the serving RRHs. 

Non-coherent transmission requires that the user implements SIC to decode its data streams from the serving RRHs, which can introduce some error propagation. To maintain mathematical tractablility, we consider a perfect SIC case, which consists the best case scenario, and thus we use the lower-bound for the rate derived in~\cite{NoncoherentCRAN8482453} to perform the resource allocation. As discussed in the coherent transmission case, and as used in~\cite{NoncoherentCRAN8482453}, this formula is based on assuming that the user has perfect knowledge of the effective precoded channel ${\bf \hat{h}}_{ru}^H {\bf \bar{w}}_{ru}$. Please refer to Remark~\ref{remark:coherentSINR} where we discuss this assumption. 
Under perfect SIC, without error propagation, a lower bound the achievable rate for the user $u$ can be defined as in~\cite{NoncoherentCRAN8482453}, which is shown in equation~\eqref{eq:rate_nonCoher}.

In a manner similar to the coherent case, we maximize the WSR subject to constraints on the power and number of users served by each RRH. Our optimization problem for the non-coherent case is written as~\eqref{eq:TotalProblem_decoupledRB}.

The solution approach is also similar to the coherent case. We can optimize the following objective function
\begin{align}\label{eq:objectiveFLag_NonCoher}
&f_4(\mathcal{W}, {\bm \gamma})
=
\sum_{u \in \mathcal{U}} \delta_{u}
\left(
\log\left( 1 + 
\gamma_{u}
\right)
- \gamma_{u}
\right)
\nonumber \\
&
+
\resizebox{0.9\columnwidth}{!}
{$\displaystyle
\sum_{u \in \mathcal{U}}
\delta_{u}
\Bigg( \frac{ \left(1 + \gamma_{u}\right) \sum_{r \in \mathcal{C}_u}
	\left|{\bf \hat{h}}_{ru}^H {\bf \bar{w}}_{ru} \right|^2 }
{ \displaystyle
	+ \sum_{r' \in \mathcal{B}} \sum_{u' \in \mathcal{E}_{r'}} {\bf \bar{w}}_{r'u'}^H \left( {\bf \hat{h}}_{r'u} {\bf \hat{h}}_{r'u}^H + \bar{\bm \Theta}_{r'u} \right) {\bf \bar{w}}_{r'u'}
	+ \sigma_z^2}
\Bigg)
$}
\end{align}
To further simplify this equation, we note that the second summation in~\eqref{eq:objectiveFLag_NonCoher} is in fact a double-sum over users ($u \in \mathcal{U}$) and RRHs ($r\in \mathcal{E}_r$). The sum is in the form  $\sum_{u \in \mathcal{U}} \left(\frac{ a_u \sum_{r \in \mathcal{C}_u} A_{ru}}{B_{u}} \right)$, which can be rewritten as $\sum_{r \in \mathcal{B}} \sum_{u \in \mathcal{E}_r} \left(\frac{a_u A_{ru}}{B_{u}} \right)$. The expression in~\eqref{eq:objectiveFLag_NonCoher} can, therefore, be rewritten as
\begin{align}
	f_5(\mathcal{W}, {\bm \gamma})
	=
	\sum_{u \in \mathcal{U}} \delta_{u}
	\left(
	\log\left( 1 + 
	\gamma_{u}
	\right)
	- \gamma_{u}
	\right)
	+ \sum_{r \in \mathcal{B}}
	f_6(r; \mathcal{W}, {\bm \gamma}),
\end{align}
where, for each RRH $r$, we have
\begin{align}\label{eq:objectiveFLag_c_term_NonCoher}
	&f_6(r; \mathcal{W}, {\bm \gamma})
	=
	\nonumber \\
	&
	\sum_{u \in \mathcal{E}_r}
	\delta_{u}
	\Bigg( \frac{ \left(1 + \gamma_{u}\right)
		\left|{\bf \hat{h}}_{ru}^H {\bf \bar{w}}_{ru} \right|^2 }
	{ \displaystyle
		\sum_{r' \in \mathcal{B}} \sum_{u' \in \mathcal{E}_{r'}} {\bf \bar{w}}_{r'u'}^H \left( {\bf \hat{h}}_{r'u} {\bf \hat{h}}_{r'u}^H + \bar{\bm \Theta}_{r'u} \right) {\bf \bar{w}}_{r'u'} + \sigma_z^2}
	\Bigg),
\end{align}
where each term in the summation in~\eqref{eq:objectiveFLag_c_term_NonCoher} is the ratio of useful signal received at user $u$ from RRH $r$ to the total power received at this user (including the useful signals). As in the coherent case, we can now use fractional programming~\cite[Corollary~1]{FR8310563} to rewrite~\eqref{eq:objectiveFLag_c_term_NonCoher} as
\begin{align}\label{eq:objectiveFLag_c_term_linearized_NonCoher}
	&f_7(r; \mathcal{W}, {\bm \gamma}, {\bm \beta}_{r})
	=
	\sum_{u \in \mathcal{E}_r}
	\Bigg(
	2 \text{Re}\left\{
	\beta_{ru}^{*}
	\sqrt{\delta_{u}\left( 1 + \gamma_{u}\right)}
	{\bf \bar{w}}_{ru}^H {\bf \hat{h}}_{ru}
	\right\}
	\nonumber \\
	&
	\resizebox{0.9\columnwidth}{!}
	{$\displaystyle
	-
	|\beta_{ru}|^2
	\left(
	\sum_{r' \in \mathcal{B}} \sum_{u' \in \mathcal{E}_{r'}} {\bf \bar{w}}_{r'u'}^H \left( {\bf \hat{h}}_{r'u} {\bf \hat{h}}_{r'u}^H + \bar{\bm \Theta}_{r'u} \right) {\bf \bar{w}}_{r'u'} + \sigma_z^2
	\right)
	\Bigg)
	$}
	,
\end{align}
where the vector ${\bm \beta}_{r} \in \mathbb{C}^{|\mathcal{E}_r| \times 1}$ includes the required auxiliary variables. The objective function for the optimization problem in~\eqref{eq:objectiveFLag_NonCoher} can therefore be written as
\begin{align}\label{eq:objectiveFLag_V2_withoutpower}
	f_8( \mathcal{W}, {\bm \gamma}, {\bm \beta})
	&=
	\sum_{u \in \mathcal{U}} \delta_{u}
	\left(
	\log\left( 1 + 
	\gamma_{u}
	\right)
	- \gamma_{u}
	\right)
	\nonumber \\
	&\quad
	+ 
	\sum_{r \in \mathcal{B}} f_7(r; \mathcal{W}, {\bm \gamma}, {\bm \beta}_{r}),
\end{align}
where ${\bm \beta} = \left[ {\bm \beta}_{1}^T \dots {\bm \beta}_{|\mathcal{B}|}^T \right]$ is the concatenation of the auxiliary variables ${\bm \beta}_{r} \in \mathbb{C}^{|\mathcal{E}_r|}$ introduced in \eqref{eq:objectiveFLag_c_term_linearized_NonCoher} for each RRH $r$. Our problem can now be reformulated as follows:
	\begin{align}\label{eq:TotalProblem_nonCoher_revised}
		\stepcounter{probNum}
		(\mathrm{P}\arabic{probNum})\quad
		\max_{ \mathcal{W}, {\bm \gamma}, {\bm \beta} }\quad & f_8( \mathcal{W}, {\bm \gamma}, {\bm \beta}) & 
		%
		\text{s.t.}\quad 
		\eqref{eq:TotalUtilityProblem_w_weights_M}, \eqref{eq:TotalUtilityProblem_w_powerB}
	\end{align}
Here~\eqref{eq:TotalUtilityProblem_w_weights_M} and~\eqref{eq:TotalUtilityProblem_w_powerB} represent constraints on the number of users served and the power at each RRH. As before, introducing Lagrange multipliers $\lambda_r$ for the $\mathrm{l}_1$ approximation to the user constraint and $\mu_r$ for the power constraint, we can obtain the optimized values of one set of variables keeping the others constant:
\vspace{-0.5em}
\begin{align}\label{eq:gamma_opt_NonCoher}
\gamma_{u} = \frac{ 
	\sum_{r \in \mathcal{C}_u}
	\left|{\bf \hat{h}}_{ru}^H {\bf \bar{w}}_{ru} \right|^2 }
{ 
\resizebox{0.9\columnwidth}{!}
{$\displaystyle
	\displaystyle
	\sum_{r' \in \mathcal{B}} \sum_{u' \in \mathcal{E}_{r'},u' \ne u} \left| {\bf \hat{h}}_{r'u}^H {\bf \bar{w}}_{r'u'} \right|^2
	+ \sum_{r' \in \mathcal{B}} \sum_{u' \in \mathcal{E}_{r'}} {\bf \bar{w}}_{r'u'}^H \bar{\bm \Theta}_{r'u} {\bf \bar{w}}_{r'u'}
	+ \sigma_z^2
$}
}
\end{align}
\vspace{-0.8em}
\begin{align}\label{eq:beta_opt_NonCoher}
\beta_{ru}
=
\frac{
	\sqrt{\delta_{u}\left( 1 + \gamma_{u}\right)}
	{\bf \bar{w}}_{ru}^H  {\bf \hat{h}}_{ru}
}
{
	\sum_{r' \in \mathcal{B}} \sum_{u' \in \mathcal{E}_{r'}} {\bf \bar{w}}_{r'u'}^H \left( {\bf \hat{h}}_{r'u} {\bf \hat{h}}_{r'u}^H + \bar{\bm \Theta}_{r'u} \right) {\bf \bar{w}}_{r'u'}
}
\end{align}
\vspace{-0.8em}
\begin{align}\label{eq:beamformer_opt_Concat_Noncoher}
&{\bf \bar{w}}_{ru}
=
\resizebox{0.88\columnwidth}{!}
{$\displaystyle
\sqrt{\delta_{u}\left( 1 + \gamma_{u}\right)}
\beta_{ru}^{*}
\Bigg( \sum_{r' \in \mathcal{B}}\sum_{u' \in \mathcal{E}_{r'}}
|\beta_{r'u'}|^2
\left( {\bf \hat{h}}_{r'u} {\bf \hat{h}}_{r'u}^H + \bar{\bm \Theta}_{r'u} \right)
$}
\nonumber \\
&\quad \quad \quad
+ \left( \mu_r + \lambda_r
\alpha_{ru}\right) {\bf I}_{M} \Bigg)^{-1}
{\bf \hat{h}}_{ru}
\end{align}
The values of $\mu_r$ and $\lambda_r$ can be determined in the same way as discussed for the coherent case.

For the case of perfect CSI, similar expressions can be written for the auxiliary variables~\eqref{eq:gamma_opt_NonCoher}, \eqref{eq:beta_opt_NonCoher}, and the beamformers~\eqref{eq:beamformer_opt_Concat_Noncoher} in the non-coherent transmission mode. The details are omitted for the sake of brevity and to avoid repetition.

\section{Numerical Results and Analysis}\label{section:results}
\begin{figure}
	\centering
	\begin{subfigure}{0.25\textwidth}
		\centering
		\includegraphics[width=0.95\textwidth]{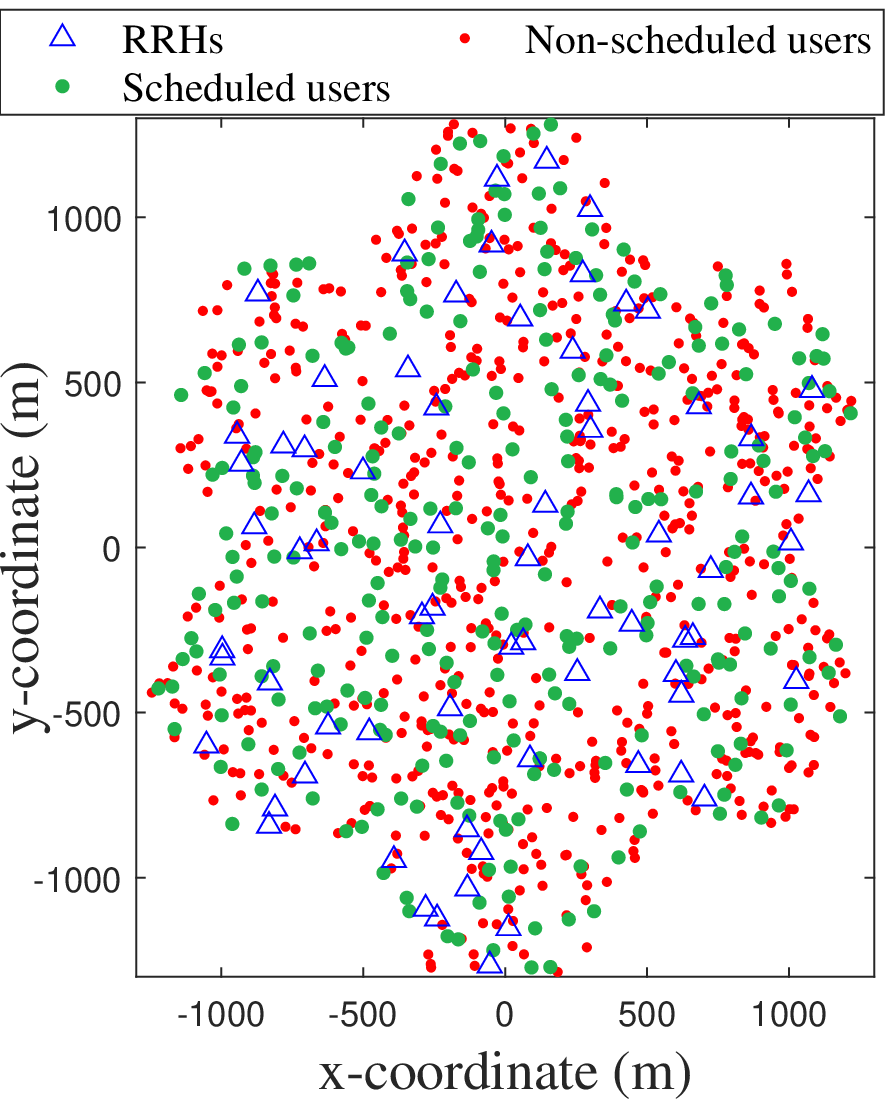}
		\captionof{figure}{Scheduled users at TS $i$.}
		\label{fig:scheduledUsers_RB10}
	\end{subfigure}%
	\begin{subfigure}{0.25\textwidth}
		\centering
		\includegraphics[width=0.95\textwidth]{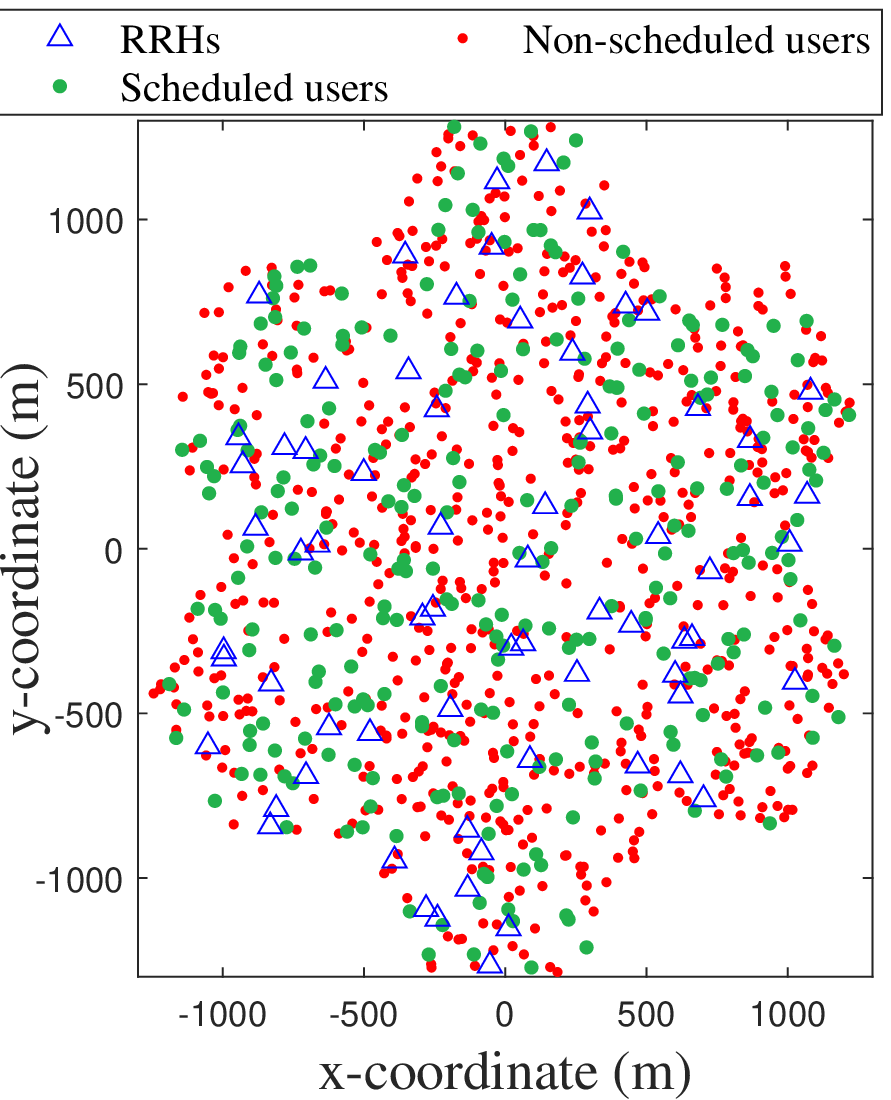}
		\captionof{figure}{Scheduled users at TS $\left(i+1\right)$.}
		\label{fig:scheduledUsers_RB11}
	\end{subfigure}%
	\vspace{-0.5em}
	\caption{Scheduled users}
	\label{fig:scheduledUsers_twoRBs}
	\vspace{-1em}
\end{figure}

We now present the results of simulations to test the efficacy of our proposed approach. To eliminate the effect of network borders, we consider a wrap-around approach comprising $Q=7$ hexagonal \emph{virtual} cells. We emphasize that the cellular structure and wrap-around is purely to cover two-dimensional space; the cells have no physical meaning as such. Each virtual cell has a radius of $500~{\rm m}$ and contains $N$ RRHs that are uniformly distributed. Similarly, users are uniformly distributed across the seven cells with a density of $\zeta_\text{users}$ users/$\text{km}^2$ but with a circular exclusion region of radius $20~{\rm m}$ around each RRH. Fig.~\ref{fig:scheduledUsers_twoRBs} illustrates a typical network. 

We use the COST231 Walfisch-Ikegami model~\cite{Walfisch14401} to define the path loss component at the $f = 1800$ MHz band as $\ell(d_{ru})\left(\mathrm{dB}\right) = - 112.4271 - 38\log_{10}\left(d_{ru}\right)$, where $d_{ru}$ is measured in $\mathrm{km}$. We average our results using Monte Carlo simulations over both network realizations and time slots (TSs), and we include the effect of the users' fairness by simulating $100$~TSs and averaging the results over the final $50$~TSs. We denote this as ``long-term results'', representing network steady state performance. We emphasize that simulating a single TS produces higher network rate than simulating many TSs, because, for a single TS, the fairness is equal for all the users, and hence users with the best channel conditions would be served. For single TS results, we do not use the label ``long-term''. 

Importantly, the long-term spectral efficiency per user is averaged over the number of the simulated TSs, thereby accounting for the fact that not every user is scheduled in all the TSs. Our configuration is static in the sense that we assume that the serving clusters $\{\mathcal{C}_u: u \in \mathcal{U}\}$ and user locations do not change during the TSs considered, but will change from realization to realization. Table~\ref{table:sim_parameters} summarizes the system parameters used (unless indicated otherwise). Importantly, in our setting, each RRH has $M = 8$  antennas.
\begin{table}[t]
	\centering
	\begin{tabular}{|p{0.3\linewidth}|p{0.22\linewidth}|p{0.32\linewidth}|}
		\hline
		\hline
		& \multicolumn{1}{l}{ \textit{\textbf{Parameter}}} & \multicolumn{1}{|l|}{\textit{\textbf{Value}}}\\
		\hline
		Cell config. & $Q$, $N$, $M$, $\zeta_\text{users}$ & $7$, $10$, $8$, $200$ users/$\text{km}^2$\\
		\hline
		Power & $p$  & $30~{\rm dBm}$\\
		\hline
		Noise spectral density, Noise figure & $S_z$, $F_z$, Bandwidth & $-174~{\rm dBm/Hz}$, $8~{\rm dBm}$, $180~{\rm KHz}$\\
		\hline
		Imperfect CSI & $\tau_d$, $(\tau_p)$, $p_u$ & $200$, $(16,\ 32,\ 64)$, $20~{\rm dBm}$\\
		\hline
		Others & 
		$\sigma_\text{shadowing}$, $\rho$, $\eta$, $\epsilon$ & 
		$4~{\rm dB}$, $\ell(0.4)$, $0.2$, $\frac{0.9p}{M}$\\
		\hline
		\hline
	\end{tabular}
	\vspace{-0.5em} 
	\caption{Simulation parameters.}
	\label{table:sim_parameters}
	\vspace{-1em}   
\end{table}

We use the weights, $\delta_u$, to impose proportional fairness., i.e., $\delta_{u}$ is inversely proportional to the achieved long-term average rate over an exponentially decaying window. Hence, at $t^\mathrm{th}$ time slot, $\delta_{u}$ is defined as~\cite{yu2011adaptive}
\begin{align} \label{eq:PropFair}
\delta_{u}^{(t)} = \frac{1}{\bar{R}_u^{(t)}}, \hspace*{0.2in} \bar{R}_u^{(t+1)} = \eta R_u^{(t)} + (1 - \eta) \bar{R}_u^{(t)}
\end{align}
where $\delta_{u}^{(t)}$ is the value of $\delta_{u}$ at time slot $t$, and $\bar{R}_u^{(t)}$ is the long-term data rate of user $u$ averaged over previous times slots, and is updated as shown with a forgetting factor $\ 0 \le \eta \le 1$. Here $R_u^{(t)}$ is the SE in TS $t$. Table~\ref{table:sim_parameters} lists our choice of $\eta = 0.2$ which, as measured using Jain's fairness index~\cite{jain1984quantitative}, provides fairness of greater than $70\%$ amongst users in $30$ allocated~TSs.

We begin by illustrating a key aspect of our algorithm. In Figs.~\ref{fig:scheduledUsers_twoRBs}(\subref{fig:scheduledUsers_RB10}) and~\ref{fig:scheduledUsers_twoRBs}(\subref{fig:scheduledUsers_RB11}), we plot the users scheduled on two different TSs, where the set of the scheduled users is dependent on the network conditions (channels, interference, etc) and most importantly, the updated user fairness. Importantly, \textit{the scheduled users} change from one TS to the next.

\subsection{Ideal CSI}\label{subsec:IdealCSI}
We start our performance analysis by comparing the performance of our framework under ideal CSI (denoted as ``$\mathrm{PI}$'' for ``Proposed-Ideal'') with three different benchmark schemes. Two of these schemes use round-robin scheduling with equal power allocation across users, one uses ZF beamforming while the other conjugate beamforming. The third scheme employs ZF with our scheduled users.

\begin{figure*}[t]
	\begin{subfigure}{.32\textwidth}
		\centering
		\includegraphics[width=0.95\linewidth]{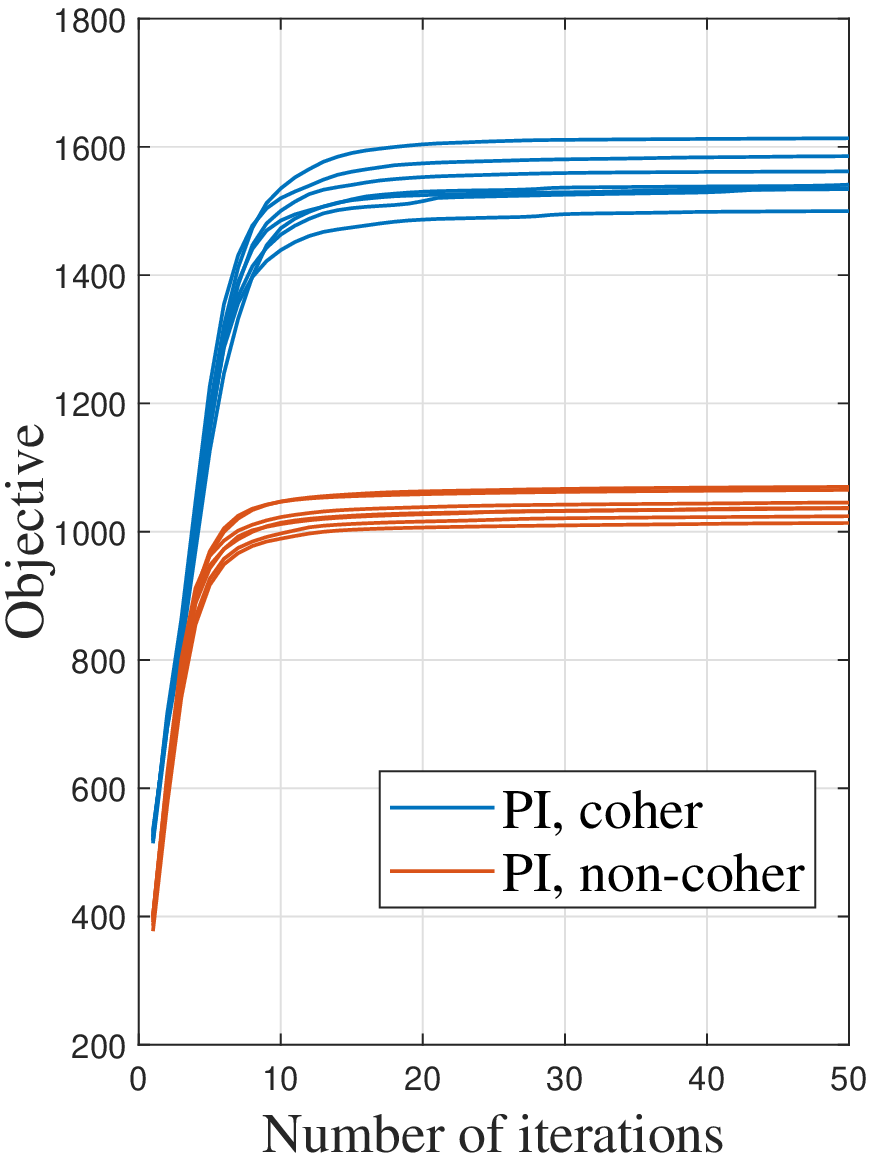}
		\caption{Convergence.}
		\label{fig:convergencePlot}
	\end{subfigure}%
	\begin{subfigure}{.32\textwidth}
		\centering
		\includegraphics[width=0.93\linewidth]{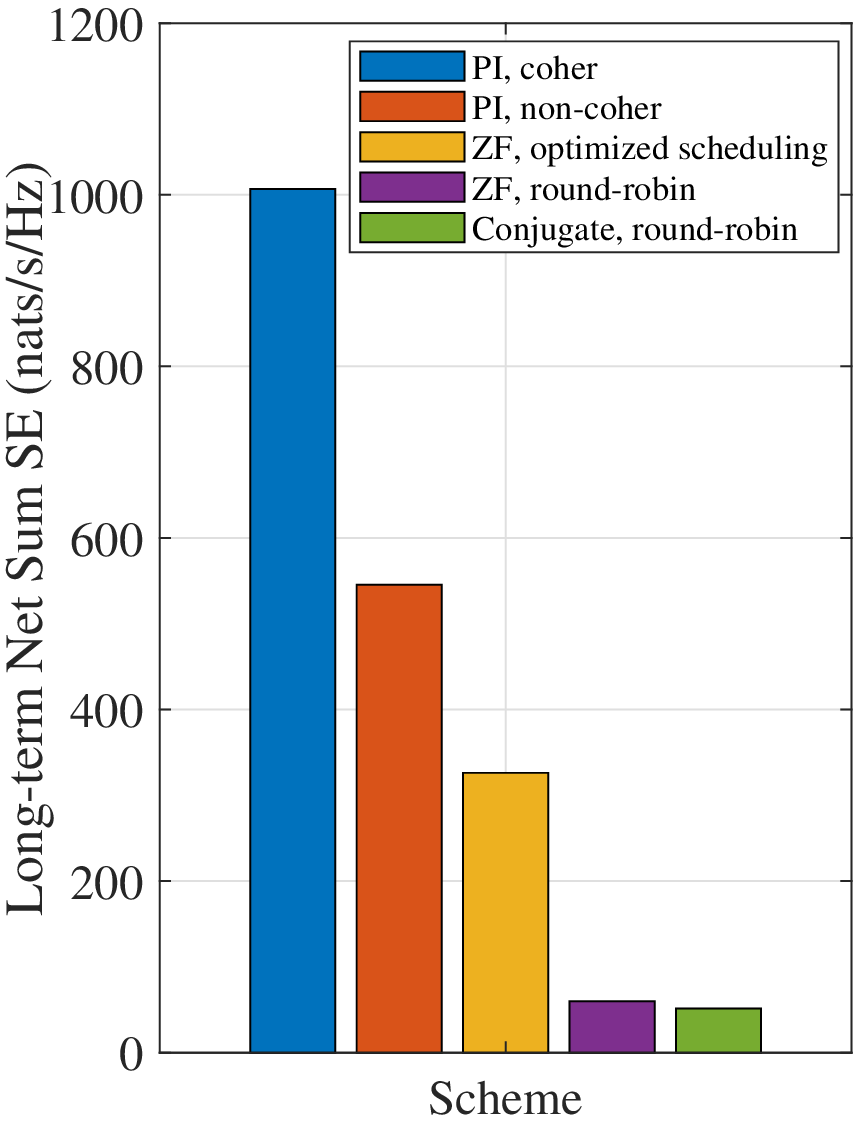}
		\vspace{0.8em}
		\caption{Long-term network sum SE.}
		\label{fig:NetSpectralEfficiency}
	\end{subfigure}%
	\begin{subfigure}{.32\textwidth}
		\centering
		\includegraphics[width=0.97\linewidth]{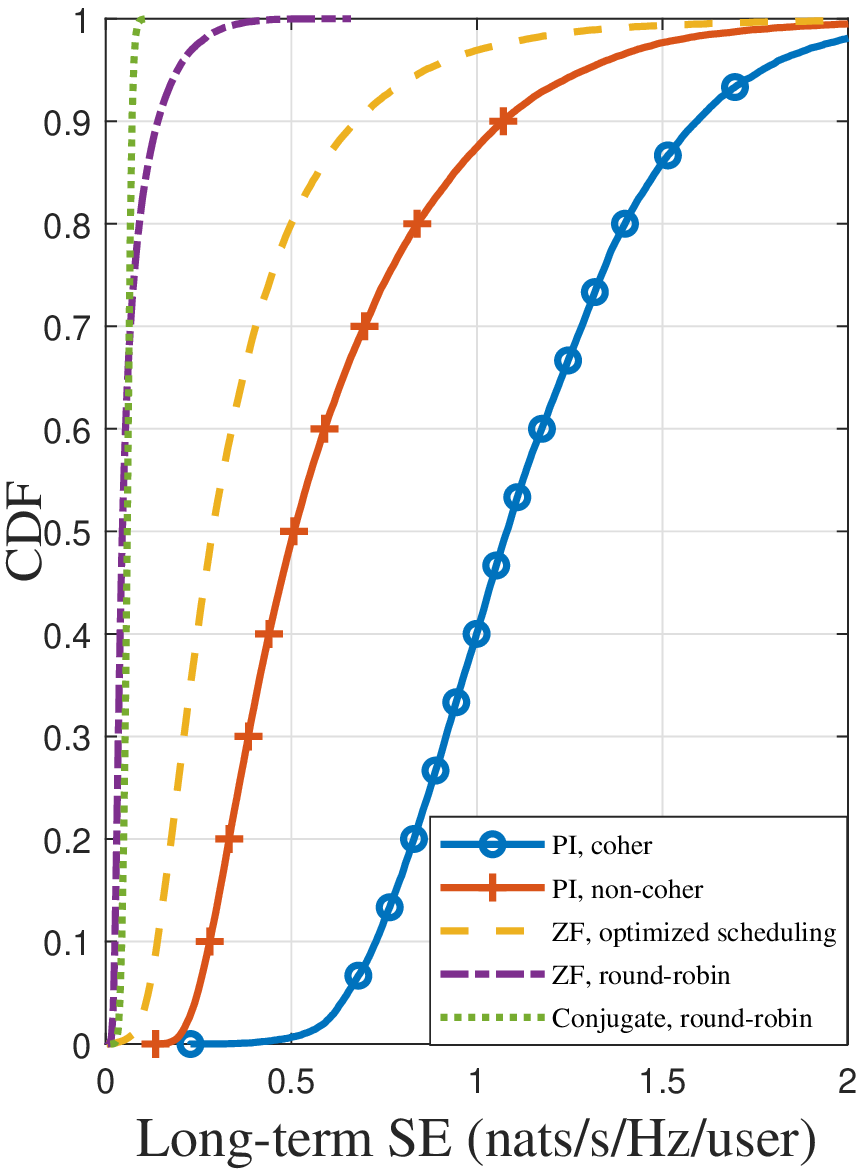}
		\caption{CDF of the long-term SE per user.}
		\label{fig:CDFspectralEfficiency}
	\end{subfigure}%
	\caption{Comparison of different schemes.}
	\vspace{-1em}
	\label{fig:IdealCSI}
\end{figure*}

In Fig.~\ref{fig:IdealCSI}(\subref{fig:convergencePlot}), we plot the network sum SE versus the number of iterations for a few representative sets of channel realizations. As is clear, in all cases, the algorithm converges smoothly for both coherent and non-coherent transmissions in a non-decreasing fashion. Furthermore, Fig.~\ref{fig:IdealCSI}(\subref{fig:NetSpectralEfficiency}) presents the average achieved long-term SE for the five cases considered. The results show a huge performance gain from using our approach compared to the ZF and  conjugate beamforming schemes with round-robin scheduling; we obtain approximately a \myResultOne-fold and \myResultTwo-fold improvement in the non-coherent transmission compared to these two schemes respectively, while we obtain $15.6$-fold and $17.1$-fold of improvement in the coherent transmission compared to the same two schemes. 

Figure~\ref{fig:IdealCSI}(\subref{fig:NetSpectralEfficiency}) also includes the network sum SE when using the ZF beamforming scheme with the optimized user scheduling obtained from our proposed approach. The figure shows that an optimized non-coherent transmission results in a \myResultThree-fold improvement compared to this optimized ZF scheme (we note that ZF also assumes coherence). This gap illustrates the importance of power allocation and beamforming. Moreover, to quantify the effect of optimized user scheduling, we compare round-robin scheduling with that of optimized scheduling for ZF beamforming. The result highlights the importance of optimizing scheduling, leading to \myResultFour-fold improvement. Finally, as expected, the coherent transmission provides a performance boost compared to the non-coherent mode; we observe a $1.84$-fold of gain in the network sum SE.

In Fig.~\ref{fig:IdealCSI}(\subref{fig:CDFspectralEfficiency}), we plot the cumulative density function (CDF) of the long-term SE of the individual users (the units are ${\rm nats/s/Hz/user}$). We see that coherent transmission provides about $2$-fold of gain in rate compared to the non-coherent mode. The figure also shows the substantial gains we can make over baseline approaches such as ZF or conjugate beamforming, especially in the $10^{\rm th}$-percentile rate (users with relatively low rates). As can be seen, the approaches based on round-robin scheduling fail to provide good spectral efficiencies to users on the long-term, further illustrating the importance of our optimization scheme. 

\subsection{Imperfect Channel State Information}
The training process uses a length-$\tau_p$ pilot sequence within every downlink data block of length $\tau_d$~\cite{9064545}. Specifically, for each user within a cluster, created using the HAC algorithm as described above, we assign one sequence from the $\tau_p \times \tau_p$ discrete Fourier Transform matrix. As listed in Table~\ref{table:sim_parameters}, we set $\tau_d = 200$ and consider the cases of $\tau_p = 16,\ 32,\ 64$. In calculating the achieved rate, we account for the pilot training overhead, i.e., $\left(\tau_d - \tau_p\right)/ \tau_d$ fraction of the time is available to transmit data. An additional time can be added if downlink training is required. To extend the HAC approach to our case with user-centric clustering, we define an \emph{area-based} pilot-reuse factor as
\vspace{-0.5em}
\begin{align}
\xi_p \triangleq \tau_p / \zeta_\text{users},
\end{align}
where, as stated at the beginning of this section, $\zeta_\text{users}$ is the density of users. 
For example, $\xi_p = 0.5$ and $\xi_p = 0.25$ implies that on-average half and one-quarter of the users, respectively, found in an area of one square-km use orthogonal pilots. For the user density specified in Table~\ref{table:sim_parameters}, the pilot sequence lengths $\tau_p = 64, 32, 16$ produce, on-average, $\xi_p = 0.32, 0.16, 0.08$ respectively.

\begin{figure*}
	\centering
	\begin{minipage}{.38\textwidth}
		\centering
		\includegraphics[width=1\linewidth]{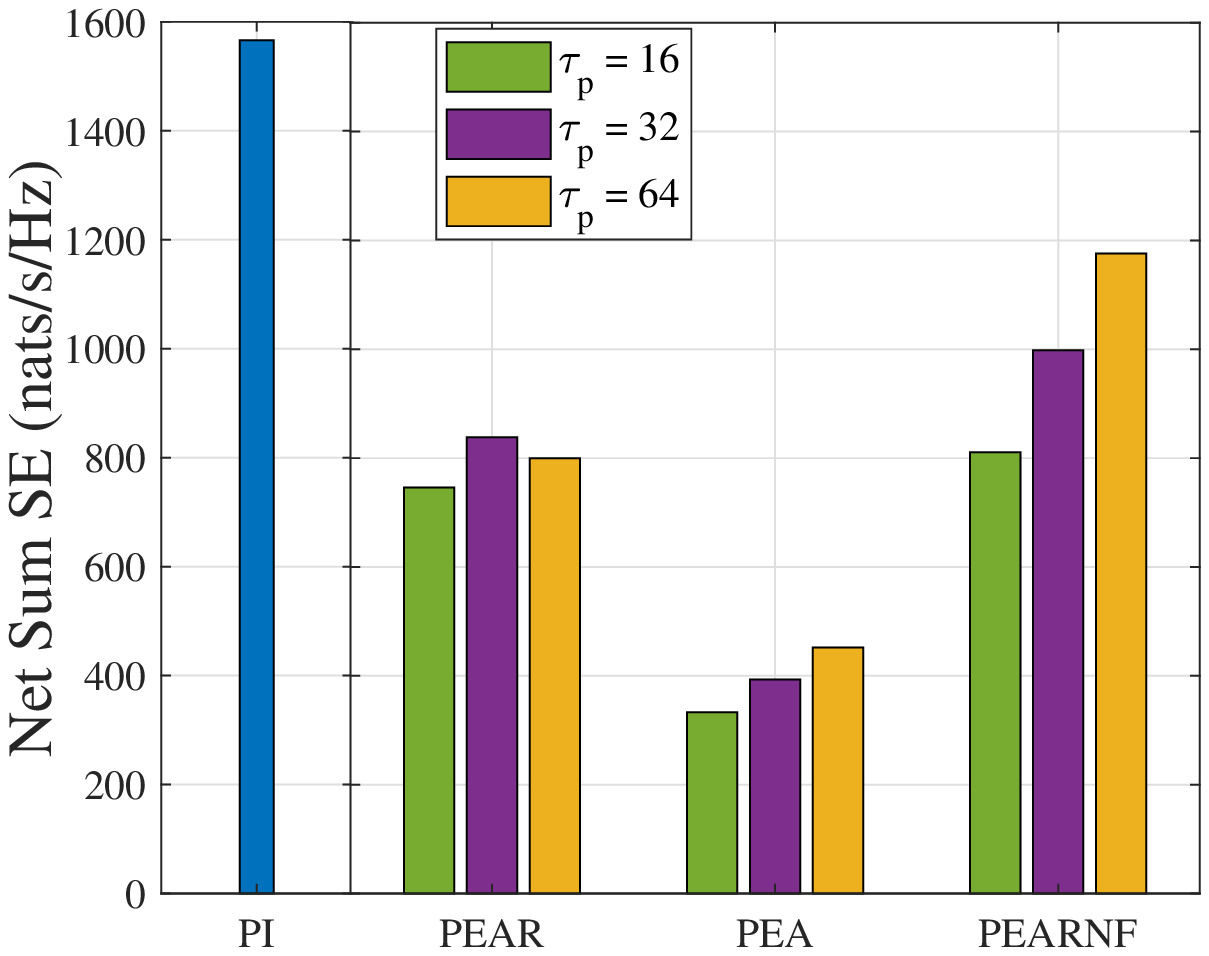}
		\caption{Sum SE with ideal (PI) and estimated CSI. Coherent mode.}
		\label{fig:coher_N5}
	\end{minipage}
	\hspace{3em}
	\begin{minipage}{.38\textwidth}
		\vspace{-0.5em}
		\centering
		\includegraphics[width=1\linewidth]{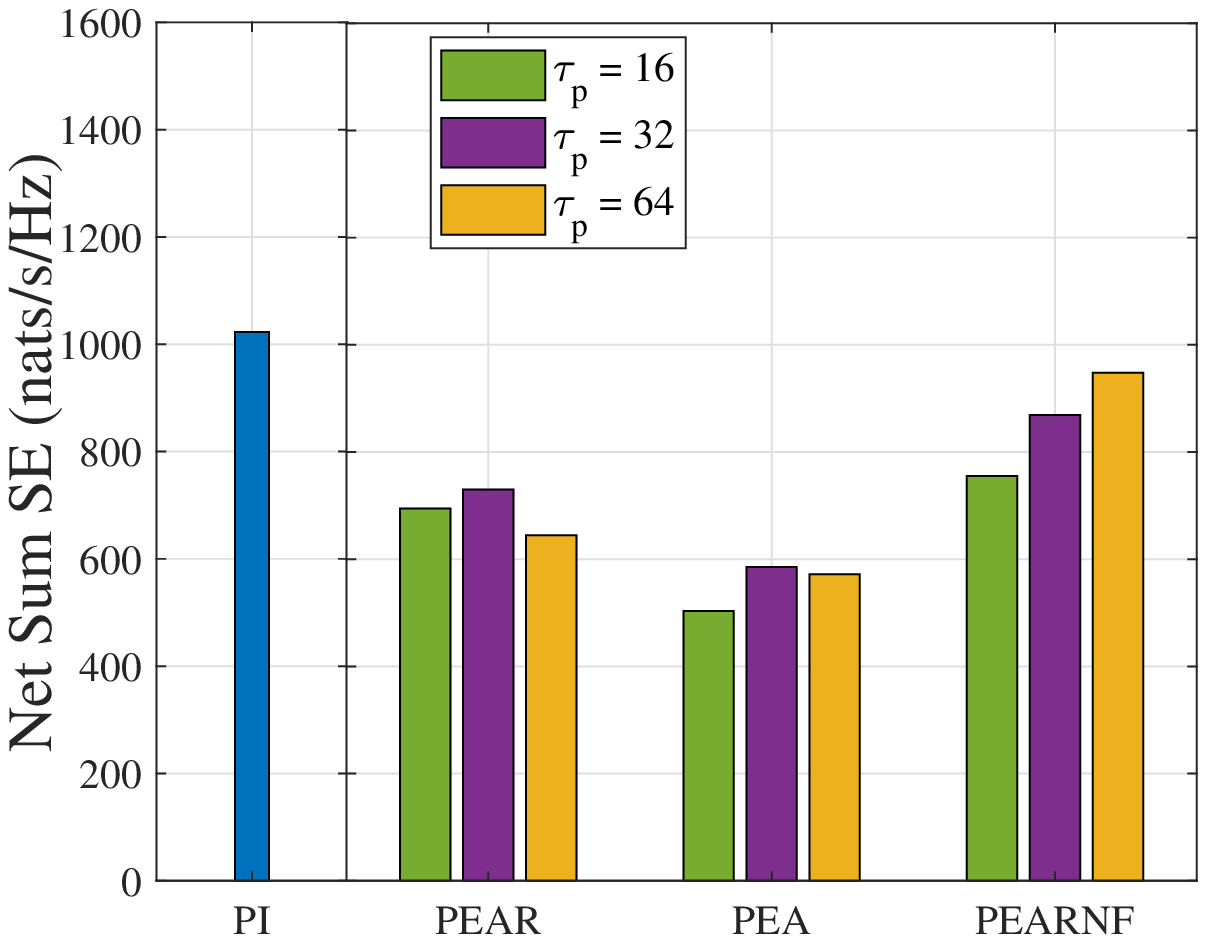}
		\caption{Sum SE with ideal (PI) and estimated CSI. Non-coherent mode.}
		\label{fig:noncoher_N5}
	\end{minipage}
\vspace{-1em}
\end{figure*}

To evaluate the performance achieved and illustrate the benefits of the proposed algorithms, we compare the following cases:
\begin{itemize}
	\item $\mathrm{PI}$ (Proposed-Ideal): Our proposed approach using \emph{ideal} channels, \textit{without} pilot-training overhead.
	\item $\mathrm{PEAR}$ (Proposed-Estimated-Actual-Robust): Our proposed approach using the \emph{estimated} channels with \emph{robust} beamforming that partially compensates for the imperfect CSI. When plotting the results, we use the \textit{actual} network rates achieved (using the optimized beams and scheduling) with the \emph{true channels} hence quantifying the actual performance achieved.
	\item $\mathrm{PEA}$ (Proposed-Estimated): To illustrate the benefits of robust BF, this result is the same as $\mathrm{PEAR}$ but without using robust beamforming, i.e. not accounting for the channel estimation error in the optimization process. Using this benchmark quantizes the  benefits of robust BF.
	\item $\mathrm{PEARNF}$ (PEAR No-Factor): Same as $\mathrm{PEAR}$, but without accounting for the pilot-training overhead, allowing us to isolate the impact of using estimated channels. 
\end{itemize}
\begin{figure*}[t]
	\centering
	\begin{minipage}{0.38\textwidth}
		\centering
		\includegraphics[width=1\linewidth]{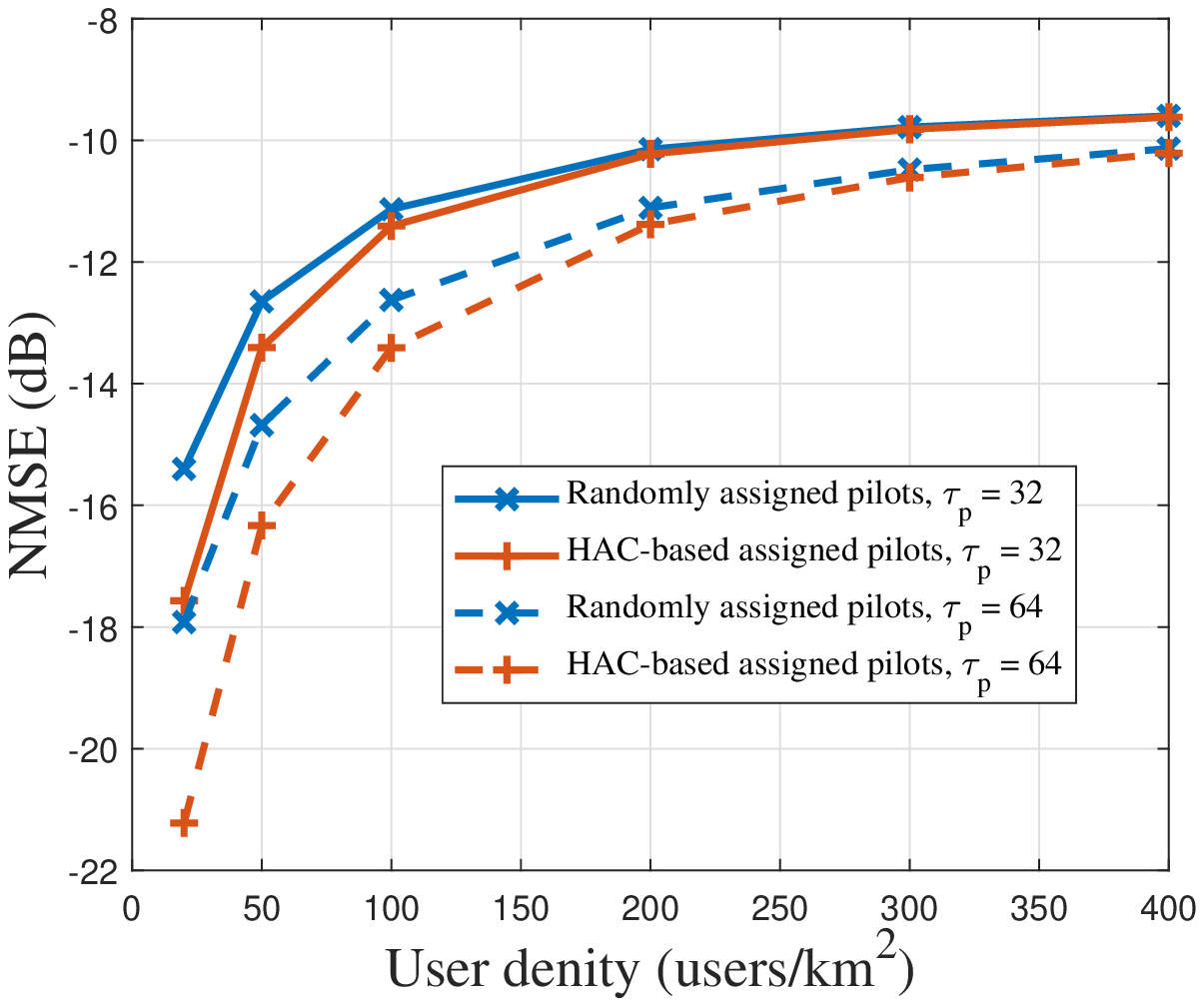}
		\caption{NMSE of estimated channels with different pilot assignments.}
		\label{fig:NMSE}
	\end{minipage}
	\hspace{3em}
	\begin{minipage}{0.38\textwidth}
		\centering
		\includegraphics[width=1\linewidth]{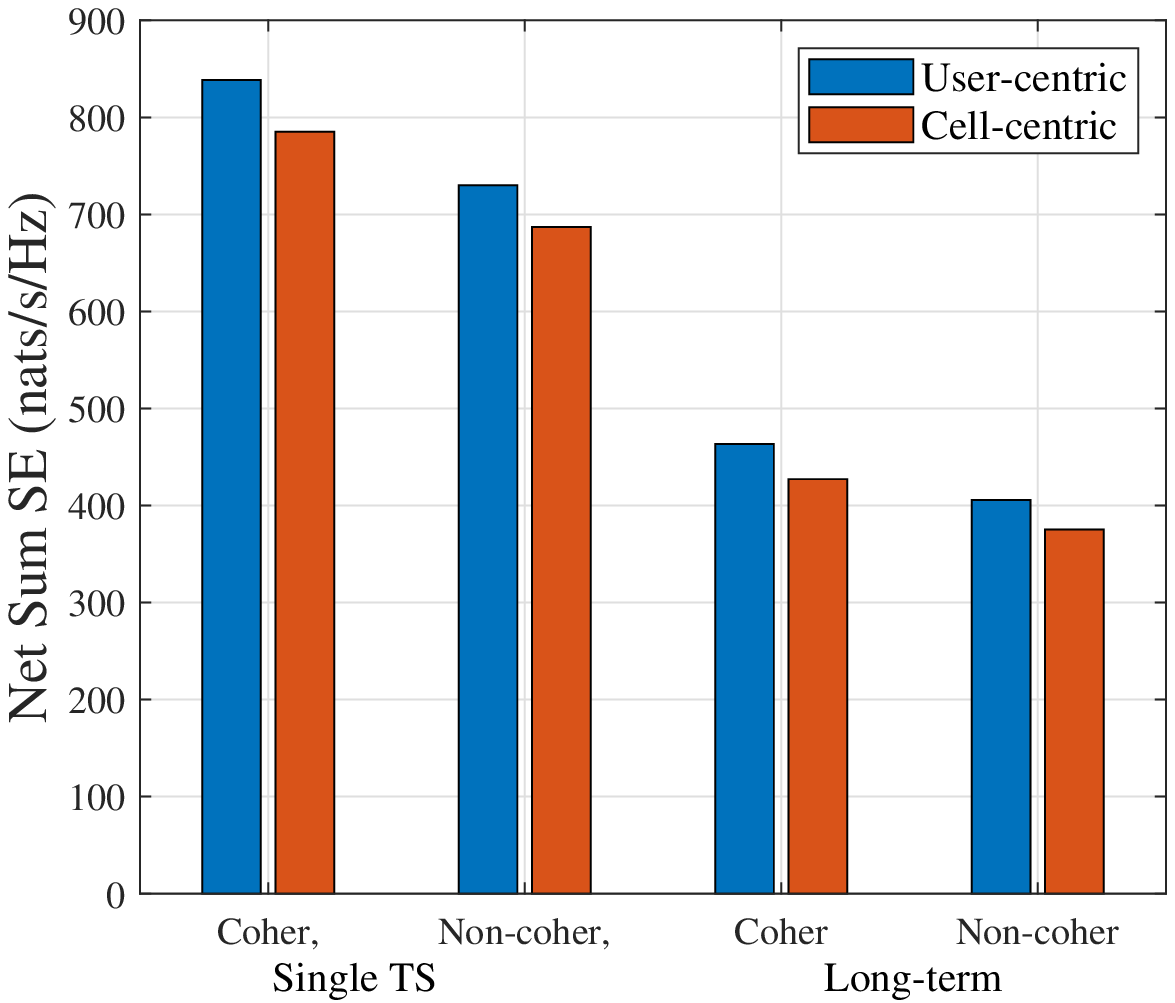}
		\caption{User-centric vs cell-centric at a single TS and long-term, $\tau_p = 32$.}
		\label{fig:UserCentric_vs_CellCentric}
	\end{minipage}
\vspace{-1em}
\end{figure*}

In Figs.~\ref{fig:coher_N5} and~\ref{fig:noncoher_N5}, we compare the achieved sum spectral efficiency using $N = 10$ RRHs per virtual cell for the coherent and non-coherent modes respectively. To obtain the sum SE, we set the weights $\delta_u = 1$. As noted earlier, all results are averaged over network realizations using Monte Carlo simulation. As with the ideal case (PI) seen in Fig.~\ref{fig:IdealCSI}, compared to the non-coherent mode, the coherent transmission provides better performance for the case of estimated CSI ($\mathrm{PEAR}$). Comparing the cases of estimated CSI, Fig.~\ref{fig:coher_N5} shows that in terms of network sum SE, for the coherent case, for pilot lengths of $\tau_p = 16, 32$ and $64$, we obtain $48.9\%$, $46.4\%$, and $52.4\%$ drop in sum SE, respectively, compared to the ideal CSI case. The figure also quantifies the gains of using robust beamforming. If we take out robust BF (i.e., PEA), we see a substantial drop in performance for all three training lengths. In addition, if we are to neglect the pilot training overhead in the imperfect CSI case (PEANF), hence isolating the effect of imperfect CSI, we would end up with a corresponding $25\%$, $36.2\%$, and $48.2\%$ drop in network performance. 

In Fig.~\ref{fig:noncoher_N5}, we plot the corresponding results for the non-coherent mode, where 
for the same pilot sequence lengths, we obtain a \myResultReuseFactorPerfRobustSingleTS$\%$, \myResultReuseFactorPerfRobustSingleTSTwo$\%$, and \myResultReuseFactorPerfRobustSingleTSThree$\%$\ drop in network sum SE, respectively, compared to the case of ideal CSI (PEAR compared to PI). These results show that the non-coherent transmission is less affected by the imperfect CSI than the coherent mode. One possible reason for this, as we will see later, could be because the coherent transmission seems to allow scheduling more users than the non-coherent mode. 

Interestingly, as seen in Figs.~\ref{fig:coher_N5} and~\ref{fig:noncoher_N5}, for both the coherent and non-coherent modes, choosing $\tau_p = 32$ provides the best \emph{average} sum SE, but only slightly better than the other cases considered. In both modes, this training sequence length provides the best balance between the quality of the estimated CSI and the reduction in time available for data transmission. 

In Fig.~\ref{fig:NMSE}, we compare the normalized mean square error (NMSE) obtained from directly assigning the pilots randomly versus the case when we use the HAC algorithm to cluster the users then again randomly assigning the pilots inside each cluster. Both results are obtained under the same network setup described in Table~\ref{table:sim_parameters} and averaged using Monte Carlo simulations. The results show that clustering the users can enhance the performance in user-centric cell-free networks with low to moderate user density because, as mentioned previously, the distance between co-pilot users is kept larger than that when directly randomly assigning the pilots.

Starting from a density of $400~{\rm users/km}^2$ the clustering does not introduce enhancements for NMSE, because the users are very close to each other. In Fig.~\ref{fig:UserCentric_vs_CellCentric}, we plot the resulted performance using our algorithm for both user-centric and cell-centric clustering. The results show that the user-centric clustering outperforms the cell-centric clustering at both for a single TS and in the long-term results.

\begin{figure*}[t]
	\centering
	\begin{subfigure}{.44\textwidth}
		\centering
		\includegraphics[width=1\linewidth]{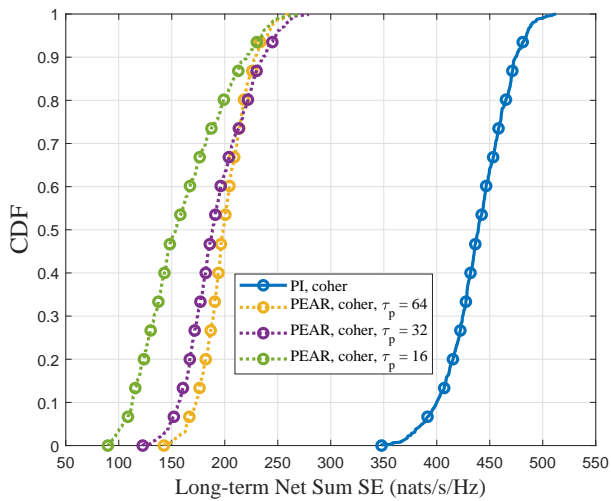}
		\caption{Coherent mode.}
		\label{sfig:CDF_netSumSE_avg_robust_coher_N5}
	\end{subfigure}%
	\hspace{3.2em}
	\begin{subfigure}{.44\textwidth}
		\centering
		\includegraphics[width=1\linewidth]{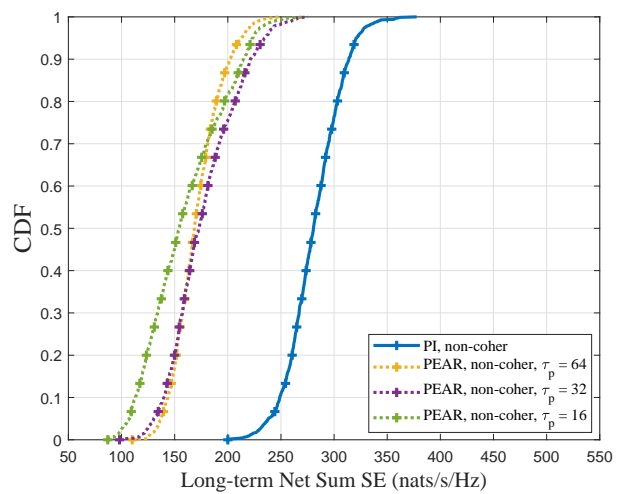}
		\caption{Non-coherent mode.}
		\label{sfig:CDF_netSumSE_avg_robust_N5}
	\end{subfigure}%
	\caption{CDF of long-term network sum SE, $N = 5$.}
	\label{fig:N5_NetSumSE}
\end{figure*}
\begin{figure*}[t]
	\centering
		\begin{subfigure}{.45\textwidth}
			\centering
			\includegraphics[width=1\linewidth]{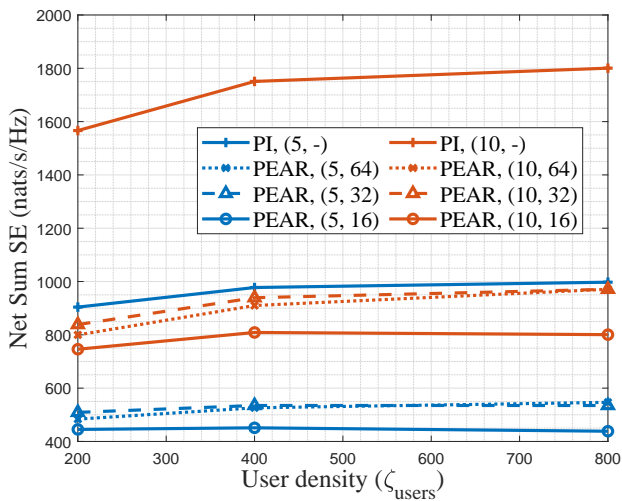}
			\caption{Coherent mode.}
			\label{fig:NetSECompare_coher}
		\end{subfigure}%
		\hspace{3em}
		\begin{subfigure}{.45\textwidth}
			\centering
			\includegraphics[width=1\linewidth]{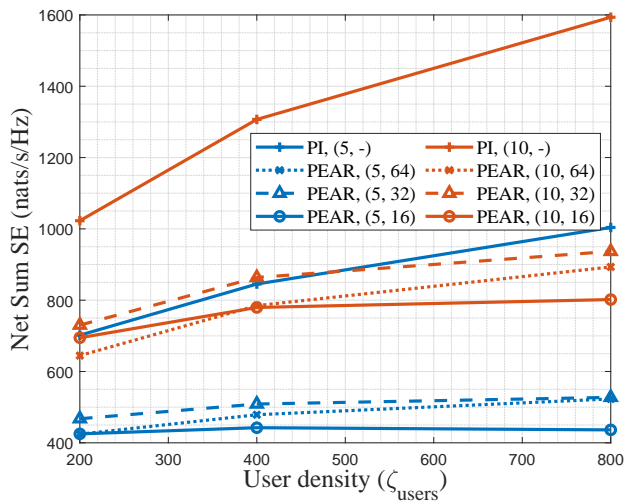}
			\caption{Non-coherent mode.}
			\label{fig:NetSECompare}
		\end{subfigure}%
		\caption{Sum of SE for different ($N$, $\tau_p$) at different densities in a single TS.}
		\label{fig:SumSEversusDensity}
\vspace{-1em}
\end{figure*}

In Fig.~\ref{fig:N5_NetSumSE}, we plot the long-term CDF of the network sum spectral efficiency for the coherent and non-coherent transmission modes for our proposed approach using $N = 5$ RRHs per virtual cell. Here, the weights evolve over time slots using the relationship in~\eqref{eq:PropFair}. This figure plots the CDF of the achieved SE for the four cases mentioned. We see a substantial drop in performance for the imperfect CSI case (PEAR) compared to that of ideal CSI (PI) due to the pilot training overhead and pilot contamination. We note that in terms of network sum SE, the results still show better efficacy for the coherent mode which provides, under the studied configuration, a $1.57$-fold, $1.28$-fold, $1.25$-fold, and $1.24$-fold gains in the long-term network sum SE compared to the non-coherent mode for the ideal and the three imperfect CSI cases respectively. 

In Fig.~\ref{fig:SumSEversusDensity}, we compare the sum SE of the network for different density of users ($\zeta_\text{users}$, in $\text{users/km}^2$) and different number of RRHs per virtual cell ($N$). The results are for the case where the fairness weights are set to unity. We summarize the following observations:
\begin{itemize}
	\item As expected, the sum SE for the network increases with the number of RRHs. This is because increasing the RRHs allows for a larger number of users scheduled from the pool of available users. 
	\item Due to user diversity, we have larger network sum SE for a higher density of users. However, after a density threshold, there is little discernible gain as this effect tapers off. We note that our user scheduling step is essential to exploit this user diversity.
	\item As before, the pilot length $\tau_p = 32$ results in the best overall performance (compared to $\tau_p = 16$ or 64), for user densities $\zeta_\text{users} = 200,\ 400,$ and $600~{\rm users}/ {\rm km}^2$. We emphasize, though, that this comment applies to the parameters of the simulation; specifically, we do not claim $\tau_p = 32$ to be a globally optimal value of training length.
\end{itemize}

\subsection{Solution Complexity}
For the coherent transmission mode, the complexity of updating ${\bm \gamma}$, ${\bm \beta}$, and ${\bm \alpha}$ is $\mathcal{O}\left(|\mathcal{U}|\right)$, $\mathcal{O}\left(|\mathcal{U}|\right)$, and $\mathcal{O}\left(|\mathcal{U}| C_\textrm{avg}\right)$ respectively, where $C_\textrm{avg}$ is the average cluster size per user, i.e., $C_\textrm{avg} = \sum_u |\mathcal{C}_u|/|\mathcal{U}|$, and it is affected by the density of users and RRHs and the large scale fading threshold for connections, $\rho$. The complexity of beamforming in Step~\ref{step:BF} in Algorithm~\ref{algortihm:w_using_weights} is $\mathcal{O}\left(|\mathcal{U}_s|^2 M^2 + |\mathcal{U}_s| M^3 \right)$, where $\mathcal{U}_s$ is the set of scheduled users~\cite{WMMSE5756489}. This leads to a worst case algorithm complexity of $\mathcal{O}\left( M^3 |\mathcal{B}|^2 + M^4 |\mathcal{B}| + |\mathcal{U}|C_\textrm{avg} + 2|\mathcal{U}|\right)$ per iteration, where the number of the scheduled users is upper bounded as $|\mathcal{U}_s| \le M |\mathcal{B}|$. 

In the non-coherent mode, the complexity for updating ${\bm \beta}$ is $\mathcal{O}\left(|\mathcal{U}| C_\textrm{avg}\right)$, because it has a dimension different from the coherent case. Hence, leading to a total algorithm complexity of at most $\mathcal{O}\left( M^3 |\mathcal{B}|^2 + M^4 |\mathcal{B}| + 2|\mathcal{U}|C_\textrm{avg} + |\mathcal{U}|\right)$ per iteration in the non-coherent mode.

\section{Conclusion}\label{section:conclusion}
This paper developed a resource allocation algorithm for user scheduling and beamforming in a distributed user-centric cell-free MIMO wireless system. In our network, individual users connect to multiple RRHs with average channel powers above a chosen threshold. Specifically, we maximized the weighted sum rate (based on a proxy for the achievable rate) while meeting constraints on the number of users each RRH serves and power. We developed an effective algorithm based on block coordinate descent, fractional programming, and compressive sensing. These steps allowed us to construct an optimization algorithm that converges smoothly in a non-decreasing fashion. One key step is user scheduling which is neglected in most of the available literature. 

In our work, we used the weights to impose proportional fairness across users. Our results show that an optimized resource allocation boosts the performance substantially compared to conventional allocation schemes especially in the long-term. Specifically, for the parameters chosen, we achieved  \myResultOne-\myResultTwo\ and $15.6$-$17.1$ fold improvements in the network long-term sum rate for non-coherent and coherent transmission respectively compared to our benchmark schemes (ZF and conjugate BF). Additionally, we analyzed the performance loss due to imperfect channel state information with robust beamforming and pilot training overhead. The high performance gain from our scheme indicates that it is important to consider optimized resource allocation and user scheduling for user-centric cell-free MIMO networks.


\bibliography{RA_UserCentric_References}
\bibliographystyle{ieeetr}

\end{document}